\documentclass[twoside]{article}
\usepackage{amssymb,amsmath}

\input BoxedEPS
\SetRokickiEPSFSpecial  
\HideDisplacementBoxes

\newcommand{\half}{{\textstyle{\frac12}}}
\newcommand{\ket}[1]{\left|#1\right\rangle}
\newcommand{\bra}[1]{\left\langle#1\right|}
\newcommand{\braket}[2]{\left\langle{#1}\mkern-2mu
\mid \mkern-2mu{#2}
\right\rangle}
\newcommand{\braketnew}[2]{\left\langle{#1}\vphantom{#2}
\mkern2mu
 \right| \mkern1mu \left.\mkern-2mu{#2}\vphantom{#1}
\right\rangle}

\let\braket\braketnew

\newcommand{\numEq}[2]{\begin{equation}
   \label{eq:#1}
   #2\end{equation}}
\newcommand{\refEq}[1]{(\ref{eq:#1})}
\newcommand{\numFig}[3]{\begin{figure}[ht]
 $$\BoxedEPSF{#3}$$
    \caption{#2}\label{fig:#1}
   \end{figure}}
\newcommand{\refFig}[1]{Fig.\thinspace\ref{fig:#1}}

\newcommand{\citeRef}[1]{\thinspace\cite{ref:#1}}
\newcommand{\eps}{\epsilon}

\newcommand{\beq}{\begin{equation}}
\newcommand{\eeq}{\end{equation}}
\newcommand{\rd}{\mathrm{d}}
\newcommand{\rP}{{\mathrm{P}}}
\newcommand{\rB}{{\mathrm{B}}}
\newcommand{\gm}{g_{\min}}
\newcommand{\ifff}{if and only if}
\newcommand{\fract}[2]{{\textstyle\frac#1#2}}

\let\epsilon\varepsilon

\let\tilde\widetilde
\let\bar\overline

 \newcommand{\Cases}[1]{\left\{ 
  \begin{array}{r@{\ ,}@{\quad}l}
  #1
  \end{array}\right.}
 \newcommand{\Casesnew}[1]{
  \begin{array}{r@{\qquad}l}
  #1
  \end{array}}

\numberwithin{equation}{section}
\begin{document}

\title{\mbox{}\vspace{-1.3in}\mbox{}\\Quantum Computation by
Adiabatic Evolution}
\markboth{Quantum Computation by Adiabatic Evolution}{E. Farhi, J.
Goldstone, S. Gutmann, and M. Sipser}

\author{Edward Farhi, Jeffrey Goldstone\thanks{farhi@mit.edu ;   
goldston@mitlns.mit.edu}\\[-0.5ex] \small
 Center for Theoretical Physics\\[-0.7ex] \small 
 Massachusetts Institute of Technology\\[-0.7ex] \small 
 Cambridge, MA  02139\\[1.5ex] Sam
Gutmann\thanks{sgutm@neu.edu}\\[-0.5ex] \small
 Department of Mathematics\\[-0.7ex] \small
  Northeastern University\\[-0.7ex] \small 
 Boston, MA 02115\\[1.5ex] Michael Sipser\thanks{sipser@math.mit.edu
\protect\newline This work was supported in part by The Department of
Energy under cooperative agreement  DE--FC02--94ER40818, by
the National Science Foundation under grant NSF 95--03322 CCR, and by a
joint NTT/LCS  research contract.}\\[-0.5ex]\small
 Department of Mathematics\\[-0.7ex] \small 
 Massachusetts Institute of Technology\\[-0.7ex] \small
 Cambridge, MA  02139\\[2ex]\small MIT CTP \# 2936\qquad
  quant-ph/0001106}

\date{}
\maketitle
\thispagestyle{empty}

\begin{abstract}\noindent  We give a quantum algorithm for solving
instances of the satisfiability problem, based on adiabatic evolution.  The
evolution of the quantum state is governed by a time-dependent
Hamiltonian that interpolates between an initial Hamiltonian, whose
ground state is easy to construct, and a final Hamiltonian, whose ground
state  encodes the satisfying assignment.  To ensure that the system
evolves to the desired final ground state, the evolution time must be big
enough.  The time required depends on the minimum energy difference
between the two lowest states of the interpolating Hamiltonian.  We are
unable to estimate this gap in general.  We give some special symmetric
cases of the satisfiability problem where the symmetry allows us to
estimate the gap and we show that, in these cases, our algorithm runs in
polynomial time.
\end{abstract}

\section{Introduction}\label{sec:0}  We present a quantum algorithm for
the satisfiability problem (and other combinatorial search problems) that
works on the principle of quantum adiabatic evolution.  

An $n$-bit instance of satisfiability is a formula
\numEq{0.1}{ C_1 \wedge C_2  \wedge \cdots \wedge C_M  }  where each
clause $C_a$ is True or False depending on the values of some subset of the
bits. For a single clause, involving only a few bits, it is easy to imagine
constructing a quantum device that evolves to a state that encodes the
satisfying assignments of the clause. The real difficulty, of course, lies in
constructing a device that produces an assignment that satisfies all
$M$~clauses. 

Our algorithm is specified by an initial state in an $n$-qubit Hilbert space
and a time-dependent Hamiltonian $H(t)$ that governs the state's evolution
according to the Schr\"odinger equation. The Hamiltonian takes the form
\numEq{0.2}{ H(t) = H_{C_1}(t) + H_{C_2}(t) + \cdots + H_{C_M}(t) } where
each $H_{C_a}$ depends only on clause $C_a$ and acts only on the bits in
$C_a$. $H(t)$ is defined for $t$ between $0$ and $T$ and is slowly varying.
The initial state, which is always the same and easy to construct, is the
ground state of $H(0)$. For each~$a$, the ground state of $H_{C_a}(T)$
encodes the satisfying assignments of clause~$C_a$. The ground state of
$H(T)$ encodes the satisfying assignments of the intersection of all the
clauses. According to the adiabatic theorem, if the evolution time~$T$ is big
enough, the state of the system at time~$T$ will be very close to the ground
state of $H(T)$, thus producing the desired solution. For this algorithm to be
considered successful we require that $T$ grow only polynomially in~$n$,
the number of bits. In this paper we analyze three examples where $T$
grows only polynomially in~$n$. We are unable to estimate the required
running time~$T$ in general.

The quantum adiabatic evolution that we are using should not be confused
with cooling. For example, simulated annealing is a classical algorithm that
attempts to find the lowest energy configuration of what we have called
$H(T)$ by generating the stochastic distribution proportional to 
$e^{-\beta H(T)}$, where $\beta$ is the inverse temperature, and gradually
lowering the temperature to zero. In contrast,  quantum adiabatic
evolution forces the state of the system to remain in the ground state of the
slowly varying~$H(t)$. 

In Section~\ref{sec:1} we present the building blocks of our algorithm in
detail. This includes some discussion of the adiabatic theorem and level
crossings. In Section~\ref{sec:2} we illustrate the method on a small
example that has three clauses, each acting on 2~bits. Each 2-bit clause has
more than one satisfying assignment but adiabatic evolution using $H(t)$ of
the form \refEq{0.2} produces the unique common satisfying assignment. In
Section~\ref{sec:3} we look at examples that grow with the number of bits
in order to study the dependence of the required running time on the
number of bits. We give three examples of 2-SAT problems, each of which
has a regular structure, which allows us to analyze the quantum evolution.
In these three cases the required evolution time~$T$ is only polynomially
big in the number of bits. We also look at a version of the Grover problem
that can be viewed as a relativized satisfiability problem. In this case our
algorithm requires exponential time to produce a solution. This had to be so,
as explained in Section~\ref{sec:grover}.  

In Section~\ref{sec:4} we show that our algorithm can be recast within the
conventional paradigm of quantum computing, involving sequences of
few-bit unitary operators.

\section{Adiabatic Evolution for Solving Satisfiability}\label{sec:1}
\setcounter{equation}{0}  In this section we present a quantum algorithm
for solving satisfiability problems.  
\subsection{The Adiabatic Theorem} A quantum system evolves according
to the Schr\"odinger equation
\numEq{1.1}{i  \frac{\rd} {\rd t}\ket{\psi(t)}= H(t)\ket{\psi(t)}} and the
adiabatic theorem\citeRef0 tells us how to follow this evolution in the case
that
$H(t)$ is slowly varying.  Consider a smooth one-parameter family of
Hamiltonians $\tilde{H}(s), 0\leq s\leq 1$, and take
\numEq{1.2}{H (t)= \tilde{H}(t/T)} so that $T$ controls the rate at which
$H(t)$ varies.  Define the instantaneous eigenstates and eigenvalues of
$\tilde{H}(s)$ by 
\numEq{1.3}{ H (s)\ket{\ell; s} = E_\ell(s)\ket{\ell; s} } with
\numEq{1.4}{ E_0(s)\leq E_1(s)\leq\cdots\leq E_{N-1}(s) } where $N$ is the
dimension of the Hilbert space.  Suppose $\ket{\psi(0)}$ is the ground state
of $\tilde{H}(0)$, that is,
\numEq{1.5}{
\ket{\psi (0)} \> =\>  \ket{\ell=0; s=0}\ . } According to the adiabatic
theorem, if the gap between the two lowest levels, $E_1(s)-E_0(s)$, is
strictly greater than zero for all $0\leq s\leq1$, then
\numEq{1.6}{\lim_{T \to\infty} \bigl| \braket{\ell=0; s=1}{\psi(T)}
\bigr|=1 \ .}

This means that the existence of a nonzero gap guarantees that
$\ket{\psi(t)}$ obeying
\refEq{1.1} remains very close to the instantaneous ground state of $H(t)$
of the form \refEq{1.2} for all
$t$ from $0$ to $T$ if $T$ is big enough.  Let us define the minimum gap by
\numEq{1.7}{
\gm =\min_{0\leq s\leq 1} \bigl(E_1(s)-E_0(s)\bigr)\ . }  A closer look at
the adiabatic theorem tells us that taking
\numEq{1.8}{ T \gg \frac{{\cal E}}{\gm^2} } where 
\numEq{1.9new}{ {\cal E} = \max_{0\le s \le 1} \Bigl| \bra{\ell =1;
s^{\mathstrut}}
\frac{\rd\tilde H }{\rd s} 
\ket{\ell=0;s^{\mathstrut}}\Bigr| } 
 can make 
\numEq{2.10}{  \bigl| \braket{\ell =0;
s=1}{\psi(T)}
\bigr| } arbitrarily close to~1.  For all of the problems that we study 
${\cal E}$ is of order a typical eigenvalue of $H$ and is not too big, so the
size of $T$ is governed by $\gm^{-2}$.

\subsection{The Satisfiability Problem} Many computationally interesting
problems can be recast into an equivalent problem of finding a variable
assignment that minimizes an ``energy" function.  As a specific example,
consider 3-SAT\null.  An $n$-bit instance of 3-SAT is a Boolean formula,
\refEq{0.1}, that is specified by a collection of Boolean clauses, each of
which involves (at most) 3 of the
$n$ bits.   Each bit $z_i$ can take the value
$0$ or
$1$ and the $i$ label runs from
$1$ to $n$.  Clause $C$ is associated with the 3 bits  labeled $i_C, j_C$, and
$k_C$.  For each clause $C$ we define an energy function
\numEq{1.9}{ h_C(z_{i_C}, z_{j_C}, z_{k_C}) = \Cases{0& \mbox{if
$(z_{i_C}, z_{j_C}, z_{k_C})$  satisfies   clause   $C$}\\ 1 & \mbox{if
$(z_{i_C}, z_{j_C}, z_{k_C})$  violates  clause $C$.}} } We then define the total
energy  $h$ as the sum of the individual
$h_C$'s,
\numEq{1.10}{ h= \sum_C h_C\ . }
 Clearly $h\geq 0$ and $h (z_1, z_2,\dots, z_n)=0$ \ifff\  $(z_1, z_2,\dots,
z_n)$ satisfies all of the clauses. Thus  finding the minimum energy
configuration of
$h$ tells us if the formula has a satisfying assignment.

We will not distinguish between conventional clauses, which compute the
OR function of each constituent variable or negated variable, and
generalized clauses, which are permitted to compute an arbitrary Boolean
function of the constituent variables.  In some of our examples it will be
more convenient to consider generalized clauses.

\subsection{The Problem  Hamiltonian \protect\boldmath$H_{\rm P}$} If
we go from classical to quantum computation we replace the bit $z_i$ by a 
spin-$\half$ qubit labeled by $\ket{z_i }$ where $z_i = 0,1$.   The states
$\ket{ z_i}$ are eigenstates of the $z$ component of the $i$-th spin,
\numEq{1.11}{\ket{0}=\Bigl(\begin{array}{c}1\\0\end{array}\Bigr)
\quad \mbox{ and }\quad
 \ket{1 } = \Bigl(\begin{array}{c}0\\1\end{array}\Bigr)} so
\numEq{1.12}{
\half (1-\sigma^{(i)}_z) \ket{z_i} = z_i\ket{z_i}  
\quad \mbox{ where }\quad 
\sigma^{(i)}_z = \Bigl(\begin{array}{cc}1 & 0\\0 & -1\end{array}\Bigr)\ . }
The Hilbert space is spanned  by the $N=2^n$ basis vectors $\ket{z_1}
\ket{z_2} 
\cdots \ket{z_n}$.  Clause $C$ is now associated with the operator
$H_{\rP,C}$, 
\numEq{1.13}{ H_{\rP,C}(\ket{z_1} \ket{z_2 } \cdots \ket{z_n}) = h_C
(z_{i_C},z_{j_C},z_{k_C})
\ket{z_1} \ket{z_2}\cdots\ket{z_n}\ . } The  Hamiltonian associated with all
of the clauses, which we call $H_\rP$,
\numEq{1.14}{ H_\rP = \sum_C  H_{\rP,C} }
 is the sum of Hamiltonians each of which acts on a fixed number of bits.
By construction, $H_\rP$ is nonnegative, that is,
$\bra{\psi}H_\rP\ket{\psi} \ge 0$ for all
$\ket{\psi}$ and $H_\rP \ket{\psi} = 0$ \ifff\  $\ket{\psi}$ is a
superposition of states of the form
$\ket{z_1}\ket{z_2}\cdots\ket{z_n}$ where $z_1,z_2,\dots,z_n$ satisfy all
of the clauses.  In this context, solving a 3-SAT problem is equivalent to
finding the ground state of a Hamiltonian. Clearly many other
computationally interesting problems can be recast in this form.

\subsection{The Initial Hamiltonian \boldmath$H_\rB$}

For a given problem, specifying $H_\rP$ is straightforward but finding its
ground state may be difficult. We now consider an $n$-bit Hamiltonian
$H_\rB$ that is also straightforward to construct but whose ground state is
simple to find. Let
 $ H_\rB^{(i)}$ be the 1-bit Hamiltonian acting on the $i$-th bit
\numEq{1.16}{
 H_\rB^{(i)} = \half (1-\sigma_x^{(i)})\quad \mbox{ with }\quad
\sigma_x^{(i)} = \Bigl(\begin{array}{cc}0 & 1\\1 & 0\end{array}\Bigr) } so
\begin{align}
 H_\rB^{(i)} \ket{x_i=x} &= x\ket{x_i=x} \nonumber\\
\intertext{where}
 \ket{x_i=0} = \frac1{\sqrt2}\Bigl(\begin{array}{c}1\\1 \end{array}\Bigr)
\quad   &\mbox{and}\quad \ket{x_i=1} =
\frac1{\sqrt2}\Bigl(\begin{array}{c}1\\-1
\end{array}\Bigr)\ . \label{eq:1.17}
\end{align} 
Continuing to take 3-SAT as our working example,
clause~$C$ is associated with the bits $i_C$,
$j_C$, and $k_C$. Now define
\numEq{2.19}{ H_{\rB, C} = H_\rB^{(i_C)} + H_\rB^{(j_C)} + H_\rB^{(k_C)}   }
and 
\numEq{2.20}{ H_\rB = \sum_C H_{\rB, C}\ . }
 The ground state of $H_\rB$ is $ \ket{x_1=0} \ket{x_2=0}\cdots
\ket{x_n=0}$. This state, written in the $z$~basis, is a superposition of all
$2^n$ basis vectors $\ket{z_1}\ket{z_2}\cdots\ket{z_n}$, 
\numEq{1.18new}{
\ket{x_1=0}\ket{x_2=0} \cdots\ket{x_n=0}=
\frac1{2^{n/2}} \sum_{z_1}\sum_{z_2}\cdots\sum_{z_n}
\ket{z_1}\ket{z_2}\cdots\ket{z_n}\ . } Note that we can also write
\numEq{2.21}{ H_\rB = \sum_{i=1}^n d_i H_\rB^{(i)} } where $d_i$ is the
number of clauses in which bit~$i$ appears in the instance of 3-SAT being
considered. 

The key feature of $H_\rB$ is that its ground state is easy to construct. The
choice we made here will lead to an $H(t)$ that is of the form \refEq{0.2},
that is, a sum of Hamiltonians associated with each clause. 

\subsection{Adiabatic Evolution} We will now use adiabatic evolution to go
from the known ground state of
$H_\rB$ to the unknown ground state of $H_\rP$. Assume for now that the
ground state of $H_\rP$ is unique. Consider
\numEq{1.18}{ H(t) = (1-t/T) H_\rB + (t/T) H_\rP } so from \refEq{1.2},
\numEq{1.19}{
\tilde H(s) = (1-s) H_\rB +  s H_\rP\ . } Prepare the system so that  it
begins at $t=0$ in the ground state of
$H(0)=H_\rB$. According to the adiabatic theorem, if $\gm$ is not zero and
the system evolves according to \refEq{1.1}, then for   $T$ big enough
$\ket{\psi(T)}$ will be very close to the ground state of
$H_\rP$, that is, the solution to the computational problem. 

Using the explicit form of \refEq{1.14} and \refEq{2.20} we see that $H(t)$
and $\tilde H(s)$ are sums of individual terms associated with each clause.
For each clause $C$ let
\numEq{2.25}{ H_C(t) = (1-t/T) H_{\rB, C} + (t/T) H_{\rP, C} } and
accordingly
\numEq{2.26}{
\tilde H_C(s) = (1-s) H_{\rB, C} + s H_{\rP, C}\ . } Then we have 
\numEq{2.27}{ H(t) = \sum_C H_C(t) } and 
\numEq{2.28}{
\tilde H(s) = \sum_C \tilde H_C(s)\ . } This gives the explicit form of $H(t)$ 
described in the Introduction  as a sum of Hamiltonians associated with
individual clauses. 

\subsection{The Size of the Minimum Gap and the\\
 Required Evolution Time} Typically $\gm$ is not zero. To see this, note
from \refEq{1.7} that vanishing $\gm$ is equivalent to there being some
value of~$s$ for which
$E_1(s) = E_0(s)$. Consider a general $2\times 2$ Hamiltonian whose
coefficients are functions of~$s$
\numEq{1.20}{
\begin{pmatrix} a(s) & c(s) + i d(s)\\ c(s) - i d(s) & b(s)
\end{pmatrix} } where $a$, $b$, $c$, and $d$ are all real. The two
eigenvalues of this matrix are equal  for some~$s$ \ifff\  $a(s)=b(s)$,  
$c(s)=0$, and $d(s)=0$. The curve $\bigl( a(s),b(s),c(s),d(s)\bigr)$ in
$\mathbb{R}^4$ will typically not intersect the line $(y,y,0,0)$ unless the
Hamiltonian has special symmetry properties.  For example, suppose the
Hamiltonian \refEq{1.20} commutes with some operator, say for
concreteness $\sigma_x$.  This implies that
$a(s)=b(s)$ and $d(s)=0$.  Now for the two eigenvalues to be equal at some
$s$ we only require $c$ to vanish at some $s$.  As $s$ varies from 0 to 1 it
would not be surprising to find $c(s)$ cross zero so we see that the
existence of a symmetry, that is, an operator which commutes with the
Hamiltonian makes level crossing more commonplace.  These arguments can
be generalized to
$N\times N$ Hamiltonians and we conclude that in the absence of
symmetry, levels  typically do not cross.  We will expand on this point
after we do some examples. 

In order for our method to be conceivably useful, it is not enough for 
$\gm$  to be nonzero. We must be sure that
$\gm$ is not so small that the evolution time~$T$ is impractically large; see
\refEq{1.8}. For an $n$-bit problem we would say that adiabatic evolution
can be used to solve the problem if $T$ is less than $n^p$ for some
fixed~$p$ whereas the method does not work if $T$ is of order $a^n$ for
some $a>1$.  Returning to \refEq{1.8} we see that the required running
time $T$ also depends on ${\cal E}$ given in \refEq{1.9new}.  Using
\refEq{1.19} we have ${\rd\tilde H }/{\rd s} =H_\rP-H_\rB$.  Therefore
${\cal E}$ can be no larger than the maximum eigenvalue of
$H_\rP-H_\rB$.  From \refEq{1.14} we see that the spectrum of $H_\rP$ is
contained in 
$\{0,1,2,\dots,M\}$ where $M$ is the number of terms in \refEq{1.14}, 
that is, the number of clauses in the problem.  From \refEq{2.21} we see
that the spectrum of $H_\rB$  is contained in 
$\{0,1,2,\dots,d\}$ where $d=\sum d_i$. For 3-SAT, $d$ is no bigger than
$3M$.  We are interested in problems for which the number of clauses
grows only as a polynomial in $n$, the number of bits.  Thus
${\cal E}$ grows at most like a polynomial in $n$ and the distinction
between polynomial and exponential running time depends entirely on
$\gm$.

We make no claims about the size of $\gm$ for any problems other than
the examples given in Section~\ref{sec:3}. We will give three examples
where $\gm$ is  of order $1/n^p$ so the evolution time~$T$ is polynomial
in~$n$.  Each of these problems has a regular structure that made
calculating
$\gm$ possible. However, the regularity of these problems also makes
them classically computationally simple. The question of whether there are
computationally difficult problems that could be solved by quantum
adiabatic evolution we must leave to future investigation. 

\subsection{The Quantum Algorithm} We have presented a general
quantum algorithm for solving SAT problems.  It consists of: 
\begin{enumerate}

\item An easily constructible initial state \refEq{1.18new}, which is the
ground state of $H_\rB$ in~\refEq{2.20}.

\item A time-dependent Hamiltonian, $H(t)$, given by \refEq{1.18} that is 
easily constructible from the given instance of the problem; see
\refEq{1.14} and~\refEq{2.20}.

\item An evolution time $T$ that also appears in \refEq{1.18}.

\item Schr\"odinger evolution according to \refEq{1.1} for time~$T$. 

\item The final state $\ket{\psi(T)}$ that for $T$ big enough will be (very
nearly) the ground state of $H_\rP$.

\item A measurement of $z_1, z_2, \dots, z_n$ in the state $\ket{\psi(T)}$.
The result of this measurement will be a satisfying assignment of formula
\refEq{0.1}, if it has one (or more). If the formula \refEq{0.1} has no
satisfying assignment, the result will still minimize the number of violated
clauses. 

\end{enumerate}

Again, the crucial question about this quantum algorithm is how big
must~$T$ be in order to solve an interesting problem.  It is not clear what
the relationship is, if any, between the required size of~$T$ and the classical
complexity of the underlying problem. The best we have been able to do is
explore examples, which is the main subject of the rest of this paper. 

\section{One-, Two-, and Three-Qubit Examples}\label{sec:2}
\setcounter{equation}{0} Here we give some one-, two-, and three-qubit
examples that illustrate some of the ideas of the introduction. 
The two-qubit examples have clauses with more than one satisfying
assignment and serve as building blocks for the three-qubit example and
 for the more complicated examples of the next
section. 

\subsection{One Qubit} Consider a one-bit problem where the single clause
is satisfied \ifff\
$z_1=1$. We then take
\numEq{2.1}{ H_\rP = \half + \half \sigma_z^{(1)} } which has
$\ket{z_1=1}$ as its ground state. For the beginning Hamiltonian we take
\refEq{2.21} with $n=1$ and $d_1=1$,
\numEq{2.2}{ H_\rB =H_\rB^{(1)} = \half - \half \sigma_x^{(1)}\ . } The
smooth interpolating Hamiltonian $\tilde H(s)$ given by \refEq{1.19} has
eigenvalues $\half(1\pm\sqrt{1-2s+2s^2})$,  which are plotted in
\refFig{1}. 
\numFig{1}{\slshape  The two eigenvalues of $\tilde{H}(s)$ for a one-qubit
example.}{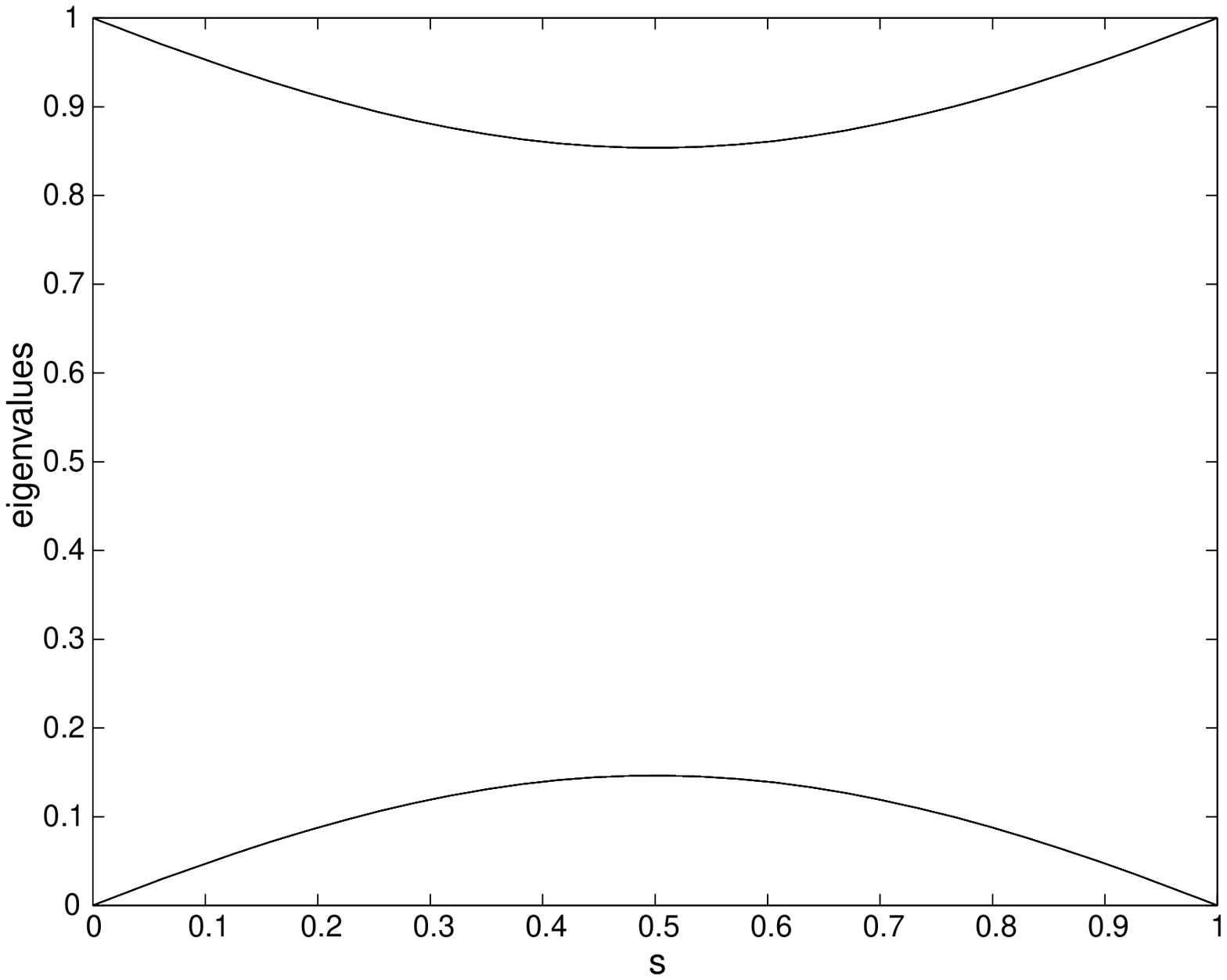 scaled 600}
\numFig{2}{\slshape  The two eigenvalues of $\tilde H(s)$ for a one-qubit
example where $H_\rB$ and $H_\rP$ are diagonal in the same basis.  The
levels cross so $\gm=0$.}{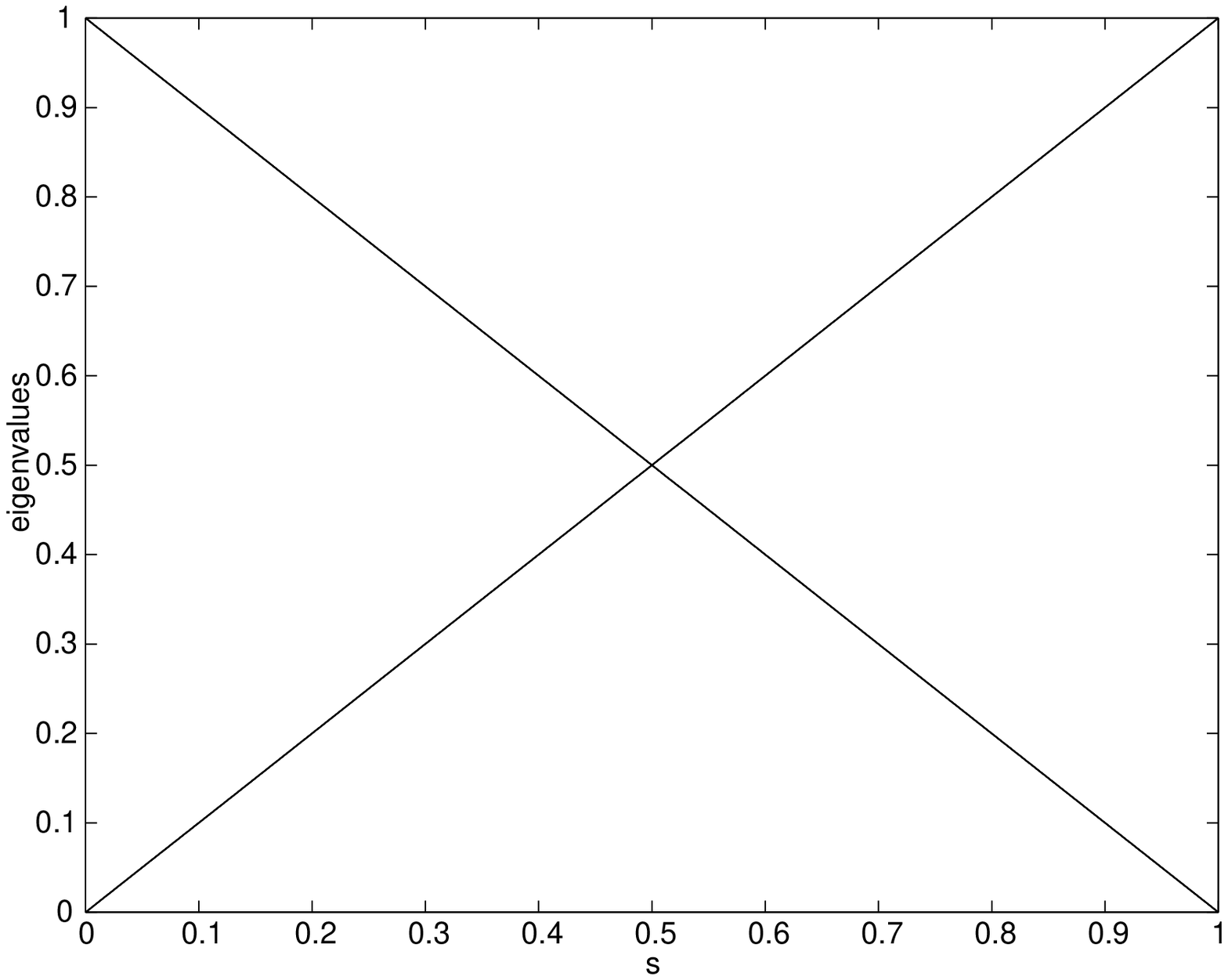 scaled 600}
  We see that $\gm$ is not
small and we could adiabatically evolve from $\ket{x_1=0}$ to
$\ket{z_1=1}$ with a modest value of~$T$.

At this point we can illustrate why we picked the beginning Hamiltonian,
$H_\rB$, to be diagonal in a basis that is \emph{not} the basis that
diagonalizes the final problem Hamiltonian~$H_\rP$. Suppose we replace
$H_\rB$ by $H'_\rB$
\numEq{2.3}{ H'_\rB = \half-\half \sigma_z^{(1)} } keeping $H_\rP$ as in
\refEq{2.1}. Now   $\tilde H(s)$ is diagonal in the
$z$-basis for all values of~$s$.  The two eigenvalues  are~$s$ and~$(1-s)$,
which are plotted in 
\refFig{2}. 
  The levels
cross so
$\gm$ is zero.  In fact there is a symmetry, $\tilde H(s)$ commutes with
$\sigma_z$ for all~$s$, so the appearance of the level cross is not
surprising.  Adiabatically evolving, starting at
$\ket{z_1=0}$, we would end up at $\ket{z_1=0}$, which is \emph{not} the
ground state of~$H_\rP$. However, if we add to $H_\rB$ any small term
that is not diagonal in the $z$ basis, we break the symmetry, and $\tilde
H(s)$ will have a nonzero gap for all~$s$. For example, the Hamiltonian
\numEq{2.4}{
\begin{bmatrix} s & {\eps}(1-s)\\[.5ex]  {\eps}(1-s) & 1-s\end{bmatrix} }
has
$\gm=\eps$ for $\eps$ small and the eigenvalues are plotted in \refFig{3}
for a small value of~$\eps$. 
This ``level repulsion'' is
typically seen in more complicated systems whereas level crossing is not. 

\numFig{3}{\slshape  A small perturbation is added to the Hamiltonian
associated with \refFig2 and we see that the levels no longer
cross.}{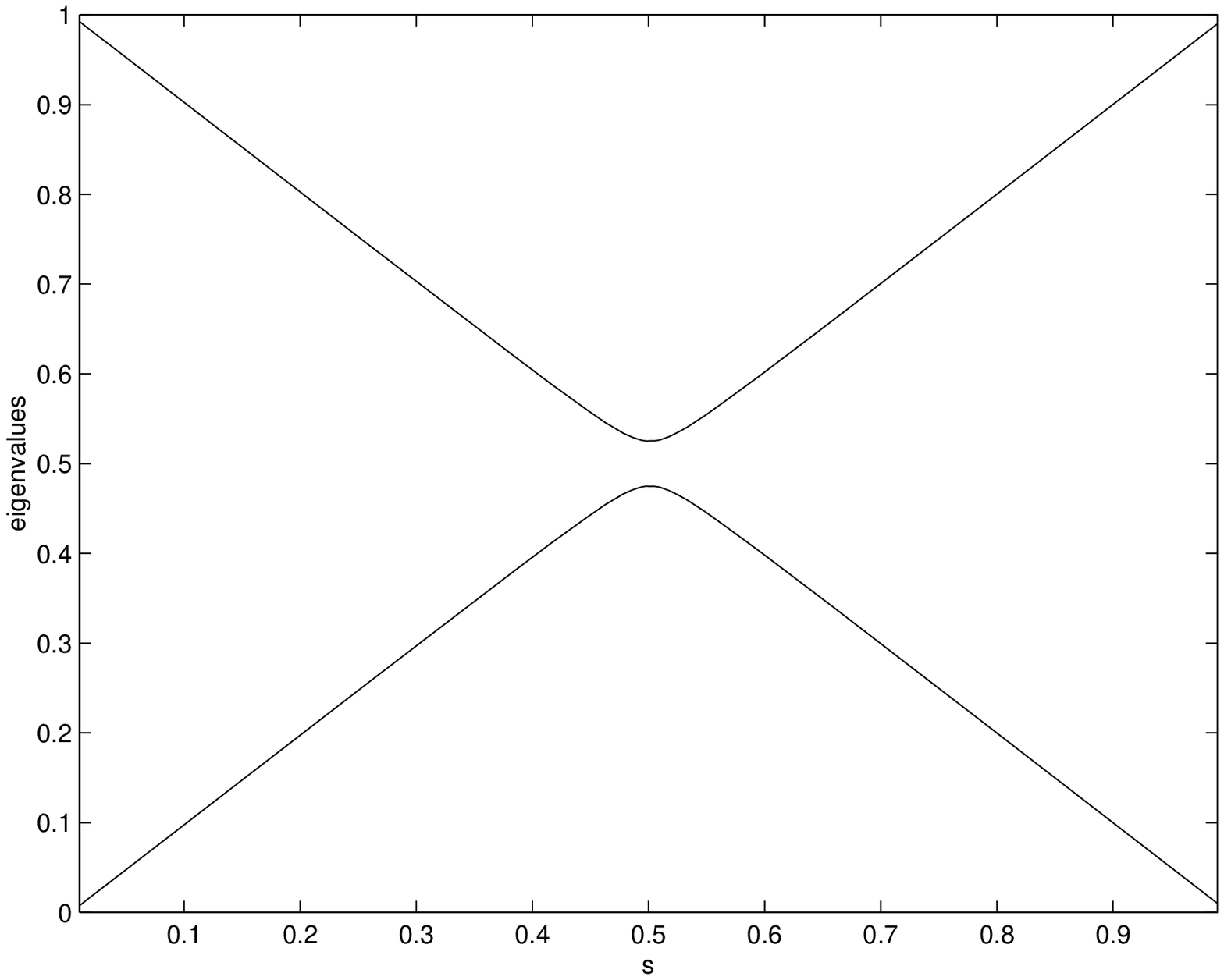 scaled 600} 

\subsection{Two Qubits} A simple two-qubit example has a single two-bit
clause that allows the bit values $01$ and
$10$ but not $00$ and $11$. We call this clause ``2-bit  disagree.'' We take
$H_\rB$ of the form \refEq{2.21} with $n=2$ and $d_1=d_2=1$,  and we
take $H_\rP$ of the form
\refEq{1.14} with the single 2-bit disagree clause.  The instantaneous
eigenvalues of $\tilde H(s)$ of the form \refEq{1.19} are shown in
\refFig{4}. There are two ground states of $H_\rP$,
\numFig{4}{\slshape  The four eigenvalues of $\tilde H(s)$ associated with
``2-bit disagree''.  The same levels are associated with ``2-bit
agree''.}{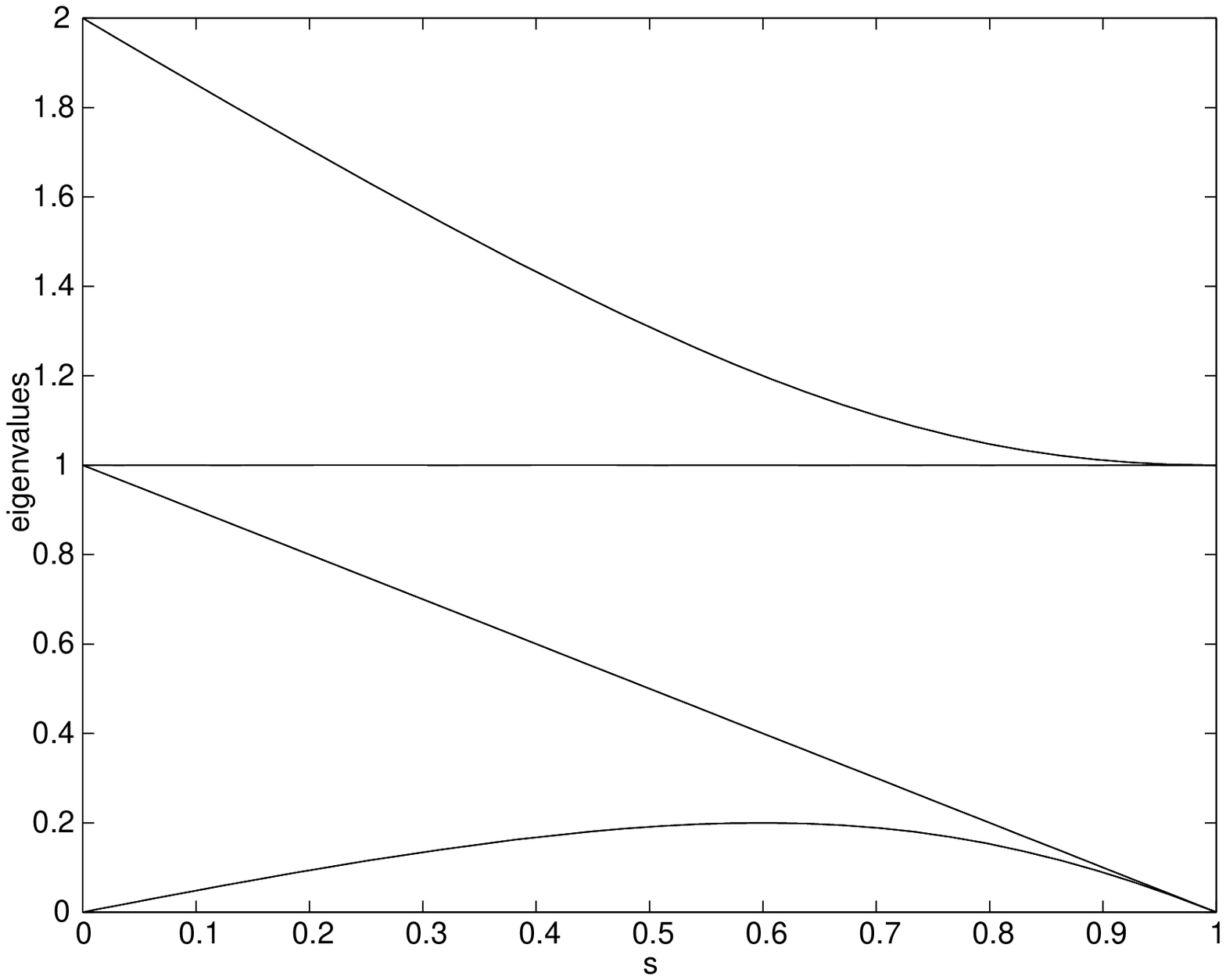 scaled 600}
$\ket{z_1=0}\ket{z_2=1}$ and $\ket{z_1=1}\ket{z_2=0}$. The starting state
$\ket{\psi(0)}$, which is the ground state of $H_\rB$, is \refEq{1.18new}
with
$n=2$. There is a bit-exchange operation $\ket{z_1}\ket{z_2} \to \ket{z_2}
\ket{z_1}$ that commutes with $\tilde H(s)$. Since the starting state
$\ket{\psi(0)}$ is invariant under the bit-exchange operation, the state
corresponding to the $s=1$ end of the lowest level in \refFig{4} is the
symmetric state $\frac1{\sqrt2} \bigl(
\ket{z_1=0}\ket{z_2=1} + \ket{z_1=1}\ket{z_2=0}\bigr)$. The next level,
$E_1(s)$, begins at the antisymmetric state 
$\frac1{\sqrt2} \bigl(
\ket{x_1=0}\ket{x_2=1} - \ket{x_1=1}\ket{x_2=0}\bigr)$ and ends at the
antisymmetric state 
$\frac1{\sqrt2} \bigl(
\ket{z_1=0}\ket{z_2=1} - \ket{z_1=1}\ket{z_2=0}\bigr)$. Because $\tilde
H(s)$ commutes with the bit-exchange operation there can be no
transitions from the symmetric to the antisymmetric states. Therefore the
$E_1(s)$ curve in \refFig{4} is irrelevant to the adiabatic evolution of the
ground state and the relevant gap is $E_2(s) - E_0(s)$. 

Closely related to  2-bit disagree is the ``2-bit agree clause,'' which has
$00$ and $11$ as satisfying assignments.  We can obtain $H_\rP$ for this
problem by taking $H_\rP$ for 2-bit disagree and acting with the operator
that takes
$\ket{z_1}\ket{z_2}\to \ket{\bar z_1}\ket{z_2}$. Note that
$H_\rB=H_\rB^{(1)} + H_\rB^{(2)}$ is invariant under this transformation as
is the starting state
$\ket{\psi(0)}$ given in
\refEq{1.18new}. This implies that the levels of $\tilde H(s)$ corresponding
to 2-bit agree are the same as those for 2-bit disagree and that beginning
with the ground state of $H_\rB$, adiabatic evolution brings you to 
$\frac1{\sqrt2} \bigl(
\ket{z_1=0}\ket{z_2=0} + \ket{z_1=1}\ket{z_2=1}\bigr)$.

Another two-bit example that we will use later is the clause ``imply''.  Here
the satisfying assignments are $00$, $01$, and $11$. The relevant level
diagram is shown in \refFig{5}. 
\numFig{5}{\slshape  The four eigenvalues of $\tilde H(s)$ associated with
the 2-bit imply clause.}{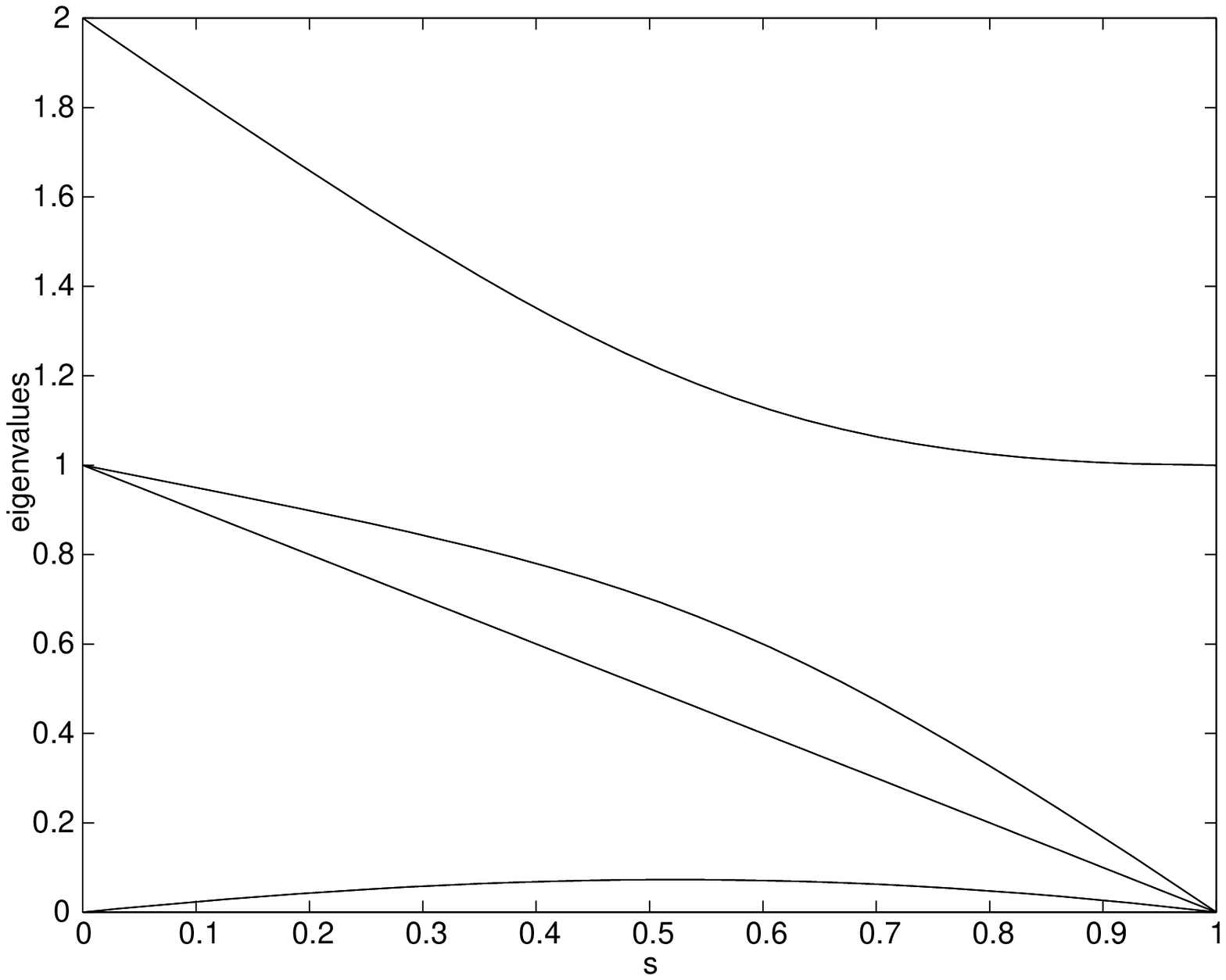 scaled 600}

\subsection{Three Qubits} Next we present a three-bit example that is built
up from two-bit clauses so we have an instance of 2-SAT with three bits.
We  take the 2-bit imply clause acting on bits 1 and~2, the 2-bit disagree
clause acting on bits 1 and~3, and the 2-bit agree clause acting on bits 2
and~3.  Although each two-bit clause has more than one satisfying
assignment, the full problem has the unique satisfying assignment~$011$. 

The corresponding quantum Hamiltonian, $\tilde H(s)=(1-s)H_\rB + s
H_\rP$,  we write as the sum of Hamiltonians each of which acts on two
bits, 
\begin{align} H_\rP &= H_{\rm imply}^{12} + H_{\rm disagree}^{13} +
H_{\rm agree}^{23}\nonumber\\[.25ex]
 H_\rB &= (H_\rB^{(1)} + H_\rB^{(2)}) + 
(H_\rB^{(1)} + H_\rB^{(3)}) +  (H_\rB^{(2)} + H_\rB^{(3)})
\ .\label{eq:3.5add}
\end{align} The eigenvalues of $\tilde H(s)$ are shown in~\refFig{6}.
\numFig{6}{\slshape  The eight levels of $\tilde H(s)$ for the 3-bit problem
with $H_\rP$ and  $H_\rB$ given by
\protect\refEq{3.5add}.}{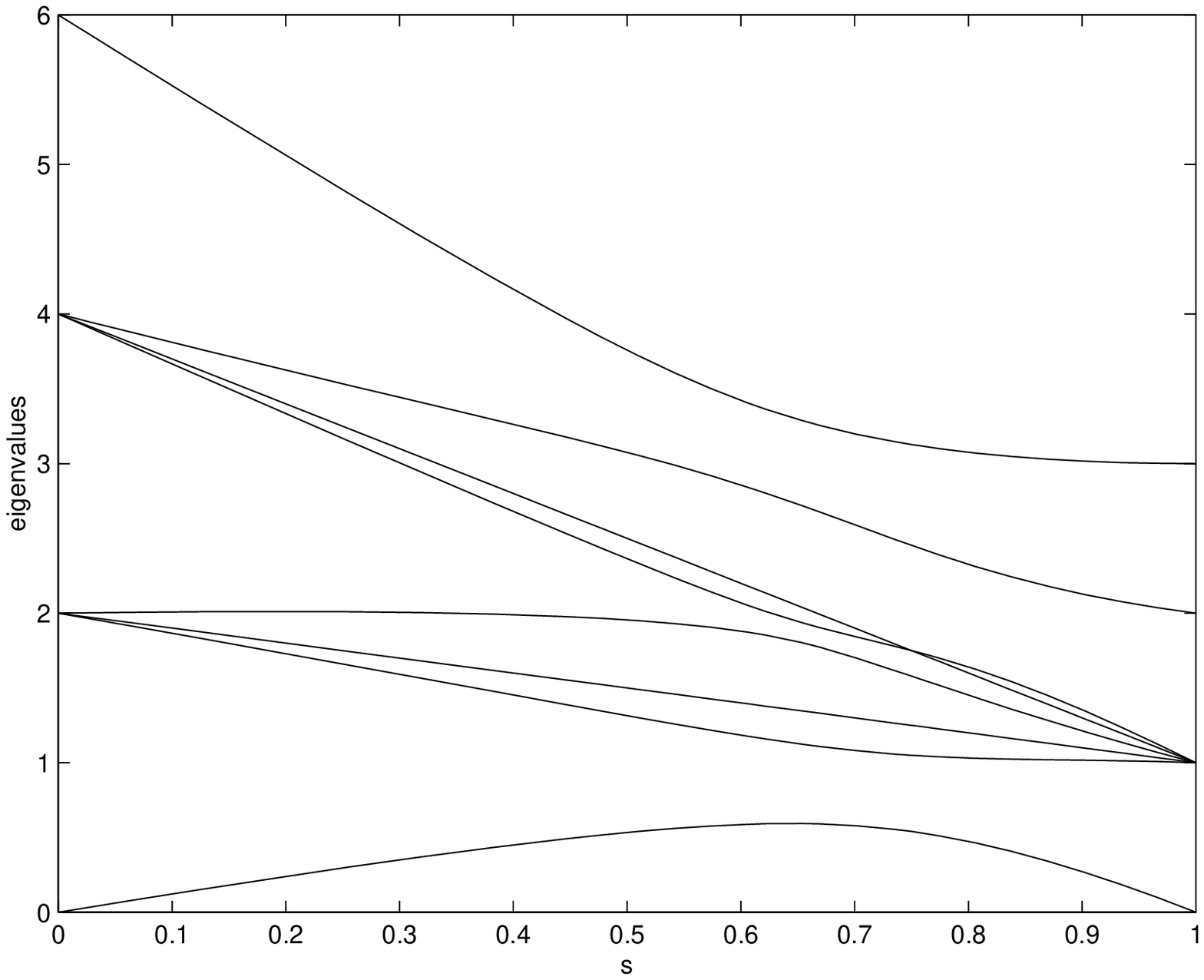 scaled 600}  We see that
$\gm$ is not zero. Starting in the ground state of $H_\rB$, and evolving
according to \refEq{1.1} with
$H(t) = \tilde H(t/T)$ the system will end up in the ground state of
$H_\rP$ for $T\gg 1/\gm^2$. This example illustrates how our algorithm
evolves to the unique satisfying assignment of several overlapping clauses
even when each separate clause has more than one satisfying assignment.

The alert reader may have noticed that two of the levels in \refFig6 cross. 
This can be understood in terms of a symmetry.  The Hamiltonian $H_\rP$
of \refEq{3.5add} is invariant under the unitary transformation
$V\bigl(\ket{z_1}\ket{z_2}\ket{z_3}\bigr) =
\ket{\bar{z}_2}\ket{\bar{z}_1}\ket{z_3}$, as is $H_\rB$.  Now the three
states with energy equal to 4 at $s=0$ are
$\ket{x_1=1}\ket{x_2=1}\ket{x_3=0}$,
$\ket{x_1=0}\ket{x_2=1}\ket{x_3=1}$, and
$\ket{x_1=1}\ket{x_2=0}\ket{x_3=1}$.  The transformation $\ket{z} \to
\ket{\bar z}$ in the $\ket{x}$ basis is $\ket{x} \to (-1)^x \ket{x}$, so the
states
$$
\ket{x_1=1}\ket{x_2=1}\ket{x_3=0}\mbox{ and }
\ket{x_1=0}\ket{x_2=1}\ket{x_3=1} - \ket{x_1=1}\ket{x_2=0}\ket{x_3=1}
$$
 are invariant under $V$, whereas
$$
\ket{x_1=0}\ket{x_2=1}\ket{x_3=1} + \ket{x_1=1}\ket{x_2=0}\ket{x_3=1}
$$ goes to minus itself.  We call these two different transformation
properties ``invariant'' and ``odd''.  Thus at $s=0$ there are two invariant
states and one odd state with energy~4.  We see from \refFig6 that one
combination of these states ends up at energy 2 when $s=1$.  The energy-2
state at $s=1$ is $\ket{z_1=0}\ket{z_2=1}\ket{z_3=0}$,
 which is invariant so the level moving across from energy 4 to energy 2 is
invariant.
 This means that one of the two levels
 that start at energy 4 and end at energy 1 is invariant and the other is
odd.   Since the Hilbert space can be decomposed into  a direct sum of the
invariant and odd subspaces and accordingly $H(t)$ is block diagonal, the
invariant and odd states
 are decoupled, and their crossing is not an unlikely occurrence.

Since, in this simple 3-bit example, we do see levels cross you may wonder
if we should expect to sometimes see the two lowest levels cross in more
complicated examples.  We now argue that we do not expect this to happen
and even if it does occur it will not effect the evolution of the ground
state.  First note that the transformation which is a symmetry of
\refEq{3.5add} is not a symmetry of the individual terms in the sum.  Thus
it is unlikely that such symmetries will typically be present in more
complicated $n$-bit examples.  However, it is inevitable that certain
instances of problems will give rise to Hamiltonians that are invariant
under some transformation.   Imagine that the transformation consists of
bit interchange and negation (in the $z$ basis) as in the example just
discussed.  Then the starting state $\ket{x=0}$ given by
\refEq{1.18new} is invariant.  Assume that $H_\rP$ has a unique ground
state $\ket{z_1 = w_1}\ket{z_2 = w_2} \cdots \ket{z_n = w_n}$.  Since
$H_\rP$ is invariant this state must transform into itself, up to a phase. 
However, from the explicit form of the ground state we see that it
transforms without a phase, that is, it is invariant.  Thus, following the
evolution of the ground state we can restrict our attention to invariant
states. The gap that matters is the smallest energy difference between the
two lowest invariant states.

\section{Examples with an Arbitrary Number of Bits}\label{sec:3}
\setcounter{equation}{0}

Here we discuss four examples  of $n$-bit instances of satisfiability. In three
of the examples the problems are classically computationally simple to solve.
These problems also have  structure that we exploit to calculate $\gm$ in
the corresponding quantum version. In each case $\gm$ goes like $1/n^p$,
so these problems can be solved in polynomial time by adiabatic quantum
evolution. The other example is the ``Grover problem''\citeRef1, which has a
single (generalized)
$n$-bit clause with a unique satisfying assignment. If we assume that we
treat the clause as an oracle, which may be queried but not analyzed, it
takes $2^n$ classical queries to find the satisfying assignment. Our
quantum version has $\gm$ of order $2^{-n/2}$, so the time required for
quantum adiabatic evolution scales like $2^n$, which means that there is
no quantum speedup.  Nonetheless, it is instructive to see how it is possible
to evaluate~$\gm$ for the Grover problem.  

\subsection{2-SAT on a Ring: Agree and Disagree}\label{sec:3.1}
\label{sec:ring} Consider an $n$-bit problem with $n$ clauses, each of
which acts only on adjacent bits,  that is, clause $C_j$ acts on bits $j$ and
$j+1$ where $j$ runs from 1 to~$n$ and bit $n+1$ is identified with
bit~$1$.  Furthermore we restrict each clause to be either ``agree'',  which
means that $00$ and $11$ are satisfying assignments or ``disagree'', which
means that
$01$ and $10$ are satisfying assignments. Suppose there are an even
number of disagree clauses so that a satisfying assignment on the ring
exists. Clearly given the list of clauses it is trivial to construct the satisfying
assignment.   Also, if
$w_1, w_2, \dots, w_n$ is a satisfying assignment, so is $\bar w_1,\bar 
w_2,
\dots,\bar  w_n$, so there are always exactly two satisfying assignments.

The quantum version of the problem has 
\numEq{3.1}{ H_\rP = H_{C_1}^{12} + H_{C_2}^{23} + \cdots + H_{C_n}^{nn+1} 
} where each $C_j$ is either agree or disagree.  The ground states of
$H_\rP$ are $\ket{w_1}\ket{w_2}\cdots\ket{w_n}$ and 
$\ket{\bar w_1}\*\ket{\bar w_2}\cdots\ket{\bar w_n}$ all in the $z$ basis.
Define the unitary transformation 
\numEq{3.2new}{
\ket{z_1}\ket{z_2}\cdots\ket{z_n} \to
\ket{z_1'}\ket{z_2'}\cdots\ket{z_n'} \qquad
\Cases{z_j'=\bar{z}_j & \mbox{if
$w_j=1$}\\[.5ex] z_j'=z_j & \mbox{if
$w_j=0$ .}} } Under this transformation $H_\rP$ becomes
\numEq{3.2}{ H_\rP = H_{\rm agree}^{12} + H_{\rm agree}^{23} + \cdots + 
H_{\rm agree}^{nn+1} } and the symmetric ground state of $H_\rP$ is 
\numEq{3.3}{
\ket w = \frac1{\sqrt2} \bigl(
\ket{z_1=0}\ket{z_2=0} \cdots \ket{z_n=0} +
\ket{z_1=1}\ket{z_2=1} \cdots \ket{z_n=1}\bigr)\ . } We take $H_\rB$ to
be \refEq{2.21} with $n$~bits and each $d_i=2$.
$H_\rB$ is invariant under the transformation just given. This implies that
the spectrum of $\tilde H(s) = (1-s) H_\rB + sH_\rP$,
 with $H_\rP$ given by \refEq{3.1}, is identical to the spectrum of $\tilde
H(s)$ 
 with $H_\rP$ given by \refEq{3.2}. Thus when we find
$\gm$ using \refEq{3.2} we will have found $\gm$ for all of the
$n$-bit agree-disagree problems initially described. 

We can write $\tilde H(s)$ using \refEq{3.2} for $H_\rP$ as 
\numEq{3.4}{
\tilde H(s) = (1-s) \sum_{j=1}^n  (1-\sigma_x^{(j)}) +  s \sum_{j=1}^n
\half(1-\sigma_z^{(j)} \sigma_z^{(j+1)})\ . } We denote the $s=0$ ground
state given by \refEq{1.18new} as $\ket{x=0}$. Define the operator~$G$
that negates the value of each bit in the $z$ basis, that is,
$G\ket{z_1}\ket{z_2}\cdots \ket{z_n} = 
\ket{\bar z_1}\ket{\bar z_2}\cdots \ket{\bar z_n}$.
 This can be written as 
\numEq{3.5}{ G = \prod_{j=1}^n \sigma_x^{(j)}\ .  } Since $G\ket{x=0} =
\ket{x=0} $ and $\bigl[ G, \tilde H(s)\bigr]=0$, we can restrict our attention
to states that are invariant under~$G$  such as~\refEq{3.3}. 

We now write \refEq{3.4} in the invariant sector as a sum of $n/2$
commuting
$2\times2$ Hamiltonians that we can diagonalize.  First we make a
standard transformation to fermion operators.  To this end we define for
$j=1,\dots,n$, 
\begin{align} b_j &= \sigma_x^{(1)} \sigma_x^{(2)}\cdots  \sigma_x^{(j-1)}
\sigma_-^{(j)} 1^{(j+1)}\cdots 1^{(n)}\nonumber\\[.5ex]
 b_j^\dagger &=
\sigma_x^{(1)} \sigma_x^{(2)}\cdots  \sigma_x^{(j-1)}
\sigma_+^{(j)} 1^{(j+1)}\cdots 1^{(n)}
\label{eq:3.6}
\end{align} where
$$
\sigma_- = \half\begin{pmatrix} 1 & -1\\ 1 & -1
\end{pmatrix} \quad \mbox{and}\quad
\sigma_+ = \half\begin{pmatrix} 1 &  1\\ -1 & -1
\end{pmatrix}\ . 
$$ It is straightforward to verify that
\begin{align}
\{ b_j, b_k\} &= 0 \nonumber\\
\{ b_j, b_k^\dagger \} &= \delta_{jk} 
\label{eq:3.7}
\end{align} where $\{ A, B\} =  AB + BA$. Furthermore
\numEq{3.8}{ b_j^\dagger b_j = \half (1-\sigma_x^{(j)}) } for $j=1,\dots,n$
and 
\numEq{3.9}{ (b_j^\dagger -  b_j) (b_{j+1}^\dagger + b_{j+1}) = 
\sigma_z^{(j)} \sigma_z^{(j+1)}  } for $j=1,\dots,n-1$. We need a bit more
care to make sense of \refEq{3.9} for $j=n$. An explicit calculation shows
that
\numEq{3.10}{ (b_n^\dagger -  b_n) (b_1^\dagger + b_1) = -G \sigma_z^{(n)}
\sigma_z^{(1)}   } where $G$  is given by \refEq{3.5}. Since we will restrict
ourselves to the $G=1$ sector, \refEq{3.9} and \refEq{3.10} are only
consistent if $b_{n+1} = - b_1$, so we take this as the definition of
$b_{n+1}$.

We can now reexpress $\tilde H(s)$ of \refEq{3.4} in terms of the $b$'s:
\numEq{3.11}{
\tilde H(s) =  \sum_{j=1}^n  \Bigl\{ 2 (1-s) b_j^\dagger b_j + \frac s2
\bigl(1- (b_j^\dagger -b_j) (b_{j+1}^\dagger + b_{j+1})\bigr)
\Bigr\}\ . } 
Because this is invariant under the translation, $b_j\to b_{j+1}$, and is
quadratic in the $b_j$ and $b_j^\dagger$, a transformation to fermion
operators associated with waves running round the ring will achieve the
desired reduction of $\tilde H(s)$.  Let
\numEq{3.12}{
\beta_p = \frac1{\sqrt n} \sum_{j=1}^n  e^{i\pi pj/n} b_j\quad \mbox{for
$p=\pm1,\pm3,\dots,\pm(n-1)$}  } which is equivalent to 
\numEq{3.13new}{ b_j = \frac1{\sqrt n}\!\!
\sum_{p=\pm1,\pm3, 
\dots,\pm (n-1)}\mkern-40mu   e^{-i\pi pj/n}
\beta_p  } and is consistent with $b_{n+1}=-b_1$. (We assume for simplicity
that $n$ is even.) Furthermore
\begin{align}
\bigl\{ \beta_p, \beta_q \bigr\} &= 0 \nonumber\\
\intertext{and}
\bigl\{ \beta_p, \beta_q^\dagger \bigr\} &= \delta_{pq}
\label{eq:3.13}
\end{align} which follows from \refEq{3.7}. Substituting \refEq{3.13new}
into \refEq{3.11} gives 
\numEq{3.15}{
\tilde H(s) =\!\!  \sum_{p=1,3,\dots,(n-1)}\mkern-30mu    A_p (s) }
where
\begin{align} A_p(s) &=2 (1-s) \bigl[
\beta_p^\dagger \beta_p + \beta_{-p}^\dagger \beta_{-p}
 \bigr]\nonumber\\
&\qquad{} + s\Bigl\{ 1-\cos\frac{\pi p}n  \bigl[
\beta_p^\dagger \beta_p - \beta_{-p} \beta^\dagger_{-p}
 \bigr]  + i \sin\frac{\pi p}n  \bigl[
\beta_{-p}^\dagger \beta_p^\dagger - \beta_p \beta_{-p}
 \bigr] \Bigr\}\ .\label{eq:3.16}
\end{align} 
The $A_p$'s commute  for different values of $p$ so we can
diagonalize each
$A_p$ separately. 

For each $p>0$ let $\ket{\Omega_p}$ be the state annihilated by both
$\beta_p$ and $\beta_{-p}$, that is, $\beta_p \ket{\Omega_p} =
\beta_{-p}\ket{\Omega_p}=0$.  When $s=0$, $\ket{\Omega_p}$ is the
ground state of $A_p$.  Now $A_p(s)$ only connects $\ket{\Omega_p}$ to
$\ket{\Sigma_p}=\beta^\dagger_{-p} \beta^\dagger_p \ket{\Omega_p}$.  In
the $\ket{\Omega_p}$, $\ket{\Sigma_p}$ basis $A_p(s)$ is
\numEq{3.17}{ A_p(s) = 
\left[ \Casesnew{s+s\cos\pi p/n  & is(\sin \pi p/n) \\[.5ex]
-is(\sin\pi p/n) & 4-3s -s\cos \pi p/n}
\right]\ . } For each $p$ the two eigenvalues of $A_p(s)$ are
\numEq{3.18}{ E^\pm_p(s) = 2-s \pm \big\{(2-3s)^2 + 4s(1-s)(1-\cos \pi
p/n)
\big\}^{\frac12}  \ . } The ground state energy of \refEq{3.15} is
$\sum\limits_p E^-_p (s)$.  The next highest energy level is $E^+_1(s) +
\sum\limits_{p=3\dots} E^-_p (s)$.  The minimum gap occurs very close to
$s=\fract23$ and is
\numEq{3.19}{
\gm \approx E^+_1(\fract23) - E^-_1(\fract23)  \approx  \frac{4\pi}{3n} }
 for $n$ large.

Referring back to \refEq{1.8} we see that the required evolution time $T$
must be much greater than ${\cal E}/\gm^2$ where for this problem ${\cal
E}$ scales like $n$ so $T \gg cn^3$ where $c$ is a constant.  We have shown
that for any set of agree and disagree clauses on an $n$-bit ring, quantum
adiabatic evolution will find the satisfying assignment in a time which
grows as a fixed power of $n$.

\subsection{The Grover Problem}
\label{sec:grover} Here we consider the Grover problem\citeRef1, which
we recast for the present context.  We have a single (generalized) clause,
$h_G$, which depends on all
$n$ bits with a unique (but unknown) satisfying assignment $w=w_1,w_2,
\dots,w_n$.  Corresponding to $h_G$ is a problem Hamiltonian
\begin{align} H_\rP \ket z &= \Cases{\ket z & z \ne w \\ 0 & z=w} 
\nonumber \\ &= 1- \ket{z=w} \bra{z=w}
\label{eq:3.20}
\end{align} where we use the shorthand  $\ket z = \ket {z_1} \ket
{z_2}\dots,\ket {z_n}$. We imagine that we can construct $H(t) = \tilde H
(t/T)$ of the form
\refEq{1.18} with $H_\rB$ given by \refEq{2.21} with $d_i=1$ for all~$i$
from~1 to~$n$.  Since we are evolving using $H(t)$ the problem is
``oracular,'' that is, we use no knowledge about the structure of $H_\rP$
which could aid us in finding $w$ other than (\ref{eq:3.20}).

We can write $\tilde H(s)$ explicitly as
\numEq{3.23}{
\tilde H(s) = (1-s) \sum^n_{j=1} \half \bigl(1 - \sigma_x^{(j)}\bigr) + s
\bigl(1 -
\ket{z=w} \bra{z=w}\bigr)  \ . } Consider the transformation given by
\refEq{3.2new}.  Under this transformation
$\tilde H(s)$ becomes
\numEq{3.24}{
\tilde H(s) = (1-s) \sum^n_{j=1} \half (1 - \sigma_x^{(j)}) + s \bigl(1 -
\ket{z=0} \bra{z=0}\bigr) \ . } Because the two Hamiltonians \refEq{3.23}
and \refEq{3.24} are unitarily equivalent they have the same spectra and
accordingly the same $\gm$.  Thus it suffices to study \refEq{3.24}.

The ground state of $\tilde H (0)$ is $\ket{x=0}$, which is symmetric under
the interchange of any two bits.  Also the operator \refEq{3.24} is
symmetric under the interchange of any two bits.  Instead of working in
the
$2^n$-dimensional space we can work in the \mbox{$(n+1)$}-dimensional
subspace of symmetrized states.  It is convenient (and perhaps more
familiar to physicists) to define these states in terms of the total spin. 
Define $\vec S = (S_x, S_y, S_z)$ by
\numEq{3.25}{ S_a = \half \sum^n_{j=1} \sigma^{(j)}_a } for $a=x,y,z$.  The
symmetrical states have $\vec{S}^{\, 2}$ equal to
$\frac{n}{2} (\frac{n}{2} +1)$, where $\vec{S}^{\, 2} = S_x^2 + S_y^2 +
S_z^2$.  We can characterize these states as either eigenstates of $S_x$ or
$S_z$
\begin{align} S_x \ket{m_x=m} = m\ket{m_x=m} \qquad m =  -\fract{n}{2},
-\fract{n}{2} +1, \dots,\fract{n}{2} \nonumber\\[.5ex]
 S_z \ket{m_z=m} =
m\ket{m_z=m} \qquad m =  -\fract{n}{2}, -\fract{n}{2} +1,
\dots,\fract{n}{2} 
\label{eq:3.26}
\end{align} where we have suppressed the total spin label since it never
changes.  In terms of the $z$ basis states previously introduced,
\numEq{3.27}{
\ket{m_z=\fract{n}{2}-k} = 
\Bigl(\begin{matrix} n \\ k \end{matrix}\Bigr)^{-\frac12}\mkern-10mu
\sum_{z_1+z_2+\cdots+ z_n=k}\mkern-30mu \ket{z_1} \ket{z_2}
\cdots \ket{z_n} } for
$k=0,1,\dots,n$.  In particular
\numEq{3.28}{
\ket{m_z=\fract{n}{2}} = \ket{z=0}\ . } Now we can write $\tilde H(s)$ in
\refEq{3.24} as
\numEq{3.29}{
\tilde H(s) = (1-s) \left[\fract{n}{2}-S_x \right] +
s\bigl[1-\ket{m_z=\fract{n}{2}}
\bra{m_z=\fract{n}{2}}\bigr] \ . } We have reduced the problem, since
$\tilde H(s)$ is now an
$(n+1)$-dimensional matrix whose elements we can simply evaluate.

We wish to solve
\numEq{3.30}{
\tilde H(s) \ket \psi = E\ket\psi } for the lowest two eigenvalues at the
value of $s$ at which they are closest.  Hitting \refEq{3.30} with
$\bra{m_x=\fract{n}{2}-r}$ we get
\begin{gather} [s+(1-s)r] \braketnew{m_x =
\fract{n}{2}-r}{\psi} -s
\braketnew{m_x=\fract{n}{2}-r}{m_z=\fract{n}{2}}\braket{m_z =
\fract{n}{2}}{\psi}\nonumber\\
 =E \braket{m_x = \fract{n}{2}-r}{\psi}  . \label{eq:3.31}
\end{gather} We replace $E$ by the variable $\lambda$ where
$E=s+(1-s) \lambda$ and obtain
\numEq{3.32}{
\frac{(1-s)}{s} \braketnew{m_x = \frac{n}{2}-r}{\psi} 
 = \frac{1}{r-\lambda}
\braketnew{m_x=\frac{n}{2}-r}{m_z=\frac{n}{2}}
\braket{m_z = \frac{n}{2}}{\psi}   .
} 
Multiply by $\braket{m_z=\frac{n}{2}}{m_x=\frac{n}{2}-r}$
and sum over $r$ to get
\numEq{3.33}{
\frac{(1-s)}{s} =\sum^n_{r=0} \frac{1}{r-\lambda} P_r } where
\numEq{3.34}{ P_r = \Big| \braket{m_z=\frac{n}{2}}{m_x=\frac{n}{2}-r}
\Big|^2 \ . }  Using \refEq{3.27} with $k=0$ and also the identical formula
with $z$ replaced by $x$ we have
\numEq{3.35}{ P_r = \frac{1}{2^n} 
\Bigl(\begin{array}{c}n\\r\end{array}\Bigr) \ . }

The eigenvalue equation \refEq{3.33} has $n+1$ roots.  By graphing the
right-hand  side of
\refEq{3.33} and keeping $0<s<1$ we see that there is one root for
$\lambda<0$, one root between 0 and 1, one root between 1 and 2, $\dots$,
and one root between $n-1$ and $n$.  The two lowest eigenvalues of
$E=s+(1-s) \lambda$ correspond to the root with $\lambda<0$ and the root
with $0<\lambda<1$.  We will now show that there is a value of $s$ for
which these two roots are both very close to zero.

The left-hand side of \refEq{3.33} ranges over all positive values as $s$
varies from 0 to 1.  Pick $s=s^*$ such that
\numEq{3.36}{
\frac{(1-s^*)}{s^*} =\sum^n_{r=1} \frac{P_r}{r} \ . } At $s=s^*$ the
eigenvalue equation \refEq{3.33} becomes
\numEq{3.37}{
\frac{P_0}{\lambda} =\sum^n_{r=1} P_r \frac{\lambda}{r(r-\lambda)}\ . }
From \refEq{3.35} we know that $P_0 =2^{-n}$.  Define $u$ by
$\lambda=2^{-n/2}u$.  Then \refEq{3.37} becomes
\numEq{3.38}{
\frac{1}{u} =\sum^n_{r=1} P_r \frac{u}{r(r-2^{-n/2}u)}\ . } Because of the
$2^{-n/2}$ we can neglect the $u$ piece in the denominator and we get
\numEq{3.39}{
\frac{1}{u^2} \approx  \sum^n_{r=1}  \frac{P_r}{r^2} } which gives
\numEq{3.40}{
\lambda \approx \pm \Big(\sum^n_{r=1}  \frac{P_r}{r^2} \Big)^{-\frac12}
2^{-n/2} } and we have
\numEq{3.41}{
\gm \approx  2(1-s^*)  \Big( \sum^n_{r=1}  \frac{P_r}{r^2} \Big)^{-\frac12}
2^{-n/2}  \ . } Now
\numEq{3.42}{
\sum^n_{r=1} \frac{P_r}{r} = \frac{2}{n} + O \Big( \frac{1}{n^2} \Big) } and
\numEq{3.42add}{
\sum^n_{r=1} \frac{P_r}{r^2} = \frac{4}{n^2} + O \Big( \frac{1}{n^3} \Big)\ . }
So using \refEq{3.36} and \refEq{3.41} we have 
\numEq{3.42add2}{
\gm \simeq 2 \cdot 2^{-\frac{n}{2}} } which is exponentially small.

In \refFig7 we show the two lowest eigenvalues of $\tilde H(s)$ for the
case of 12 bits.  If you evolve too quickly the system jumps across the gap
and you do not end up in the ground state of $\tilde H(1)$.

\numFig{7}{\slshape  The two lowest eigenvalues of $\tilde H(s)$ for the
Grover problem with 12 bits.}{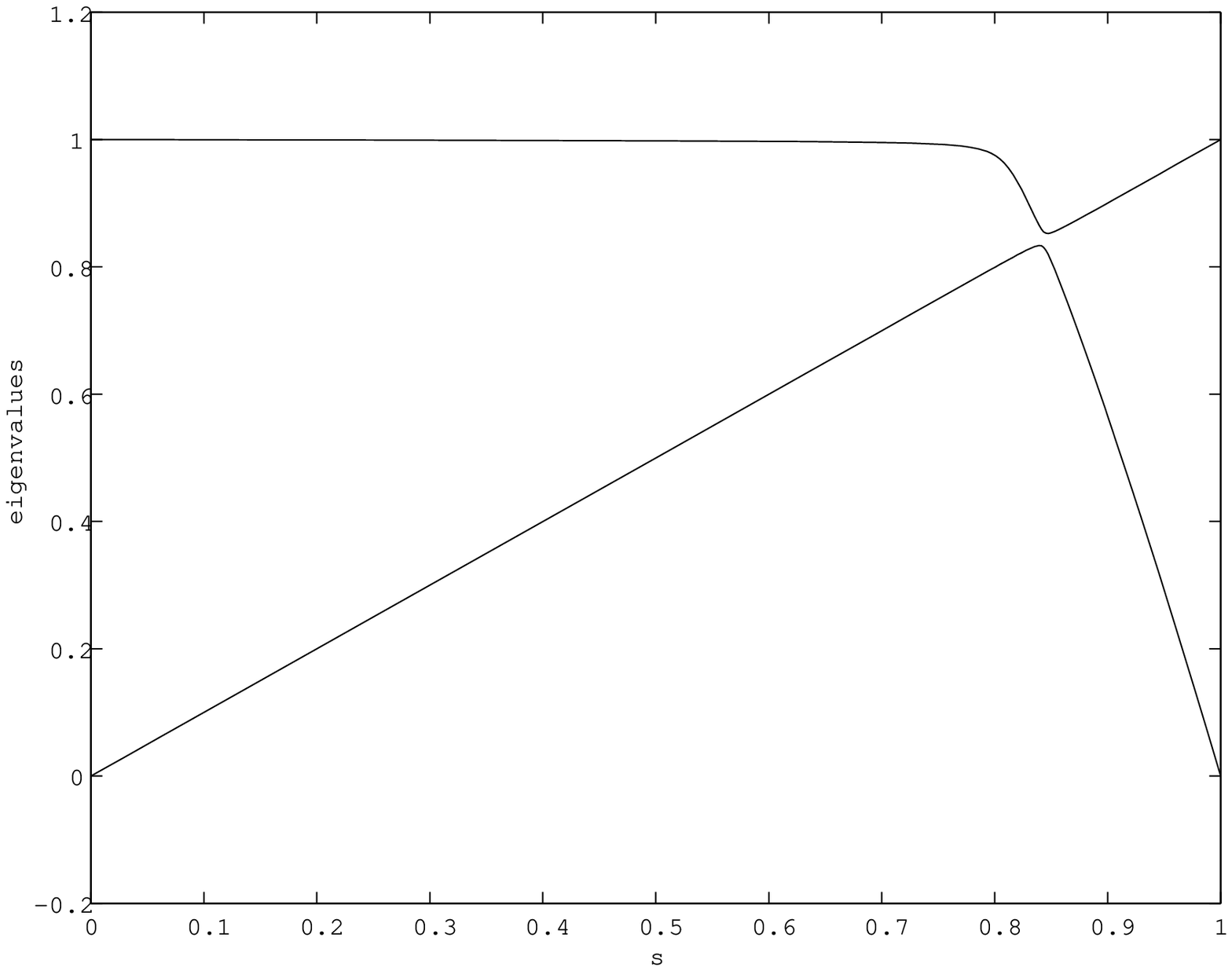 scaled 600}

That $\gm$ goes like $2^{-n/2}$ means that the required time for finding
the satisfying assignment grows like $2^n$ and quantum adiabatic
evolution is doing no better than the classical algorithm which checks all
$2^n$ variable assignments.  In reference~\citeRef2 a Hamiltonian version
of the Grover problem was studied with a time dependent Hamiltonian of
the form
\numEq{3.43}{ H(t) = H_D (t) + (1-\ket{z=w} \bra{z=w}) \ . } The goal was
to choose $H_D(t)$ without knowing $w$ so that Schr\"odinger evolution
from a $w$-independent initial state would bring the system to
$\ket{z=w}$ in time $T$.  There it was shown how to choose $H_D$ so that
the required running time $T$ grows as $2^{\frac n2}$, which is then
interpreted as the square-root speedup found by Grover.  It was also
shown that for any $H_D(t)$, $T$ must be at least of order
$2^{\frac n2}$ for  the quantum evolution to succeed for all $w$.  (The
continuous time bound found in\citeRef2 is closely related to the query
bound found first in\citeRef3.)
 A slight modification of the argument which gives this lower bound can be
made for quantum evolution with
\numEq{3.44}{ H(t) =H_D(t)  + \frac{t}{T} (1-\ket{z=w} \bra{z=w})  } and
again $T$ must be at least of order $2^{\frac{n}{2}}$.  The adiabatic
evolution we studied in this section corresponds to $H_D(t)=(1-t/T)H_\rB$
with $H_\rB$ as described above.  The lower bound just discussed shows
that no choice of $H_\rB$ can achieve better than square-root speedup.

\subsection{The Bush of Implications}\label{sec:3.3}
\label{sec:bush} Ultimately we would like to know if there are general (and
identifiable) features of problems which can tell us about the size of
$\gm$.  For the 2-SAT example of Section~\ref{sec:3.1}, $\gm$ is of order
$1/n$ whereas for the Grover problem it is of order $2^{-n/2}$.  In the
Grover case $H_\rP$ has the property that $2^n-1$ states have energy 1,
that is, there are an exponential number of states  just above the ground
state.  For the ring problem this is not so.  With $H_\rP$ of the form
\refEq{3.2} there are no states with energy 1 and (roughly) $n^2$ states
with energy 2.  Here we present an example with an exponential number
of states with energy 1 but for which the gap is of order
$1/n^p$.  This tells us that we cannot judge the size of the minimum gap
just from knowledge of the degeneracy of the first level above the ground
state of
$H_\rP$. 

The example we consider has $n+1$ bits labeled $0,1,2\dots,n$.  There are
$n$ 2-bit imply clauses, each of which involves bit 0 and one of the other
$n$ bits.  Recall that the imply clause is satisfied by the bit values $00,01$
and 11 but not by 10.  Furthermore we have a one-bit clause that is
satisfied only if bit 0 has the value 1.  The unique satisfying assignment of
all clauses is
$z_0=1, z_1=1, z_2=1, \dots,z_n=1$.

\numFig{8}{\slshape  The bush of implications. There is a one-bit clause
that  is satisfied if bit~0 has a value of~1. There are $n$ imply clauses. The
$j^{\rm th}$ imply clause is satisfied unless bit~0 has the value~1 and
bit~$j$ has the value~0.}{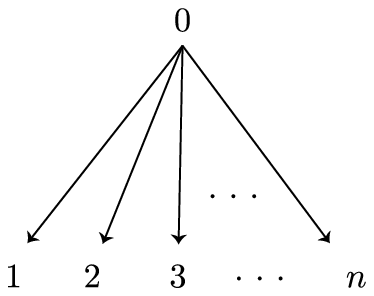}

Suppose that $z_0=0$.  Any of the $2^n$ values of $z_1, z_2, \dots,z_n$
satisfy all of the imply clauses.  Only the one bit clause is not satisfied, so
these $2^n$ variable assignments violate only one clause.  There are $n$
other variable assignments that violate only one clause.  These have all bits
set to 1 except for the $k$th bit where $1\le k\le n$.  In total there are
$2^n+n$ assignments that violate only one clause and accordingly there are
an exponential number of states with energy~1.

We can write $H_\rP$ explicitly as
\numEq{3.45}{ H_\rP=\half (1+\sigma_z^{(0)}) + {\textstyle \frac{1}{4}}
\sum^n_{j=1} (1-\sigma_z^{(0)})(1+\sigma_z^{(j)}) \ . } To evaluate $H_\rB$
from \refEq{2.21} note that bit~0 is involved in $n+1$ clauses whereas
bits~1 through~$n$ are each involved in only one clause, so
\numEq{4.47}{ H_\rB = (n+1)\half (1-\sigma_x^{(0)}) + \sum_{i=1}^n \half 
(1-\sigma_x^{(i)})\ . }
 Then $\tilde H(s)$ in terms of the spin operators
\refEq{3.25} is
\numEq{3.46}{
\tilde H(s)=(1-s) \Big[ \frac{n+1}2 (1-\sigma_x^{(0)}) + \frac{n}{2} -S_x
\Big]  + s \Big[ \half (1+\sigma_z^{(0)}) + \half (1-\sigma_z^{(0)})
\Big(\frac{n}{2} +S_z\Big)
\Big]\ . 
}
 We need only consider states that are symmetrized in the bits 1
to $n$.  We can label the relevant states as $\ket{z_0}\ket{m_z}$ where
$z_0$ gives the value of bit 0 and $m_z$ labels the $z$ component of the
total spin as in
\refEq{3.26}.  We need to know the matrix elements of $S_x$ in the
$\ket{m_z}$ basis.  These are
\begin{align}
\left\langle{m'_z}\mkern-2mu
\mid S_x \mkern-2mu \mid \mkern-2mu{m_z}
\right\rangle 
 &= \half \Big[\Big(\frac{n}{2} \Big(\frac{n}{2}+1\Big) - m^2_z - m_z
\Big)^{\frac12}\delta_{m_z, m'_z-1}\nonumber\\
&\qquad {} +
\Big(\frac{n}{2} \Big(\frac{n}{2}+1\Big) - m^{\prime 2}_z - m'_z
\Big)^{\frac12}\delta_{m'_z, m_z-1} \Big] \ .\label{eq:3.47} 
\end{align}

\numFig{9}{\slshape  The two lowest eigenvalues of $\tilde H(s)$ for the
bush of implications with $n=50$. The visible gap indicates that $\gm$ is not
exponentially small.}{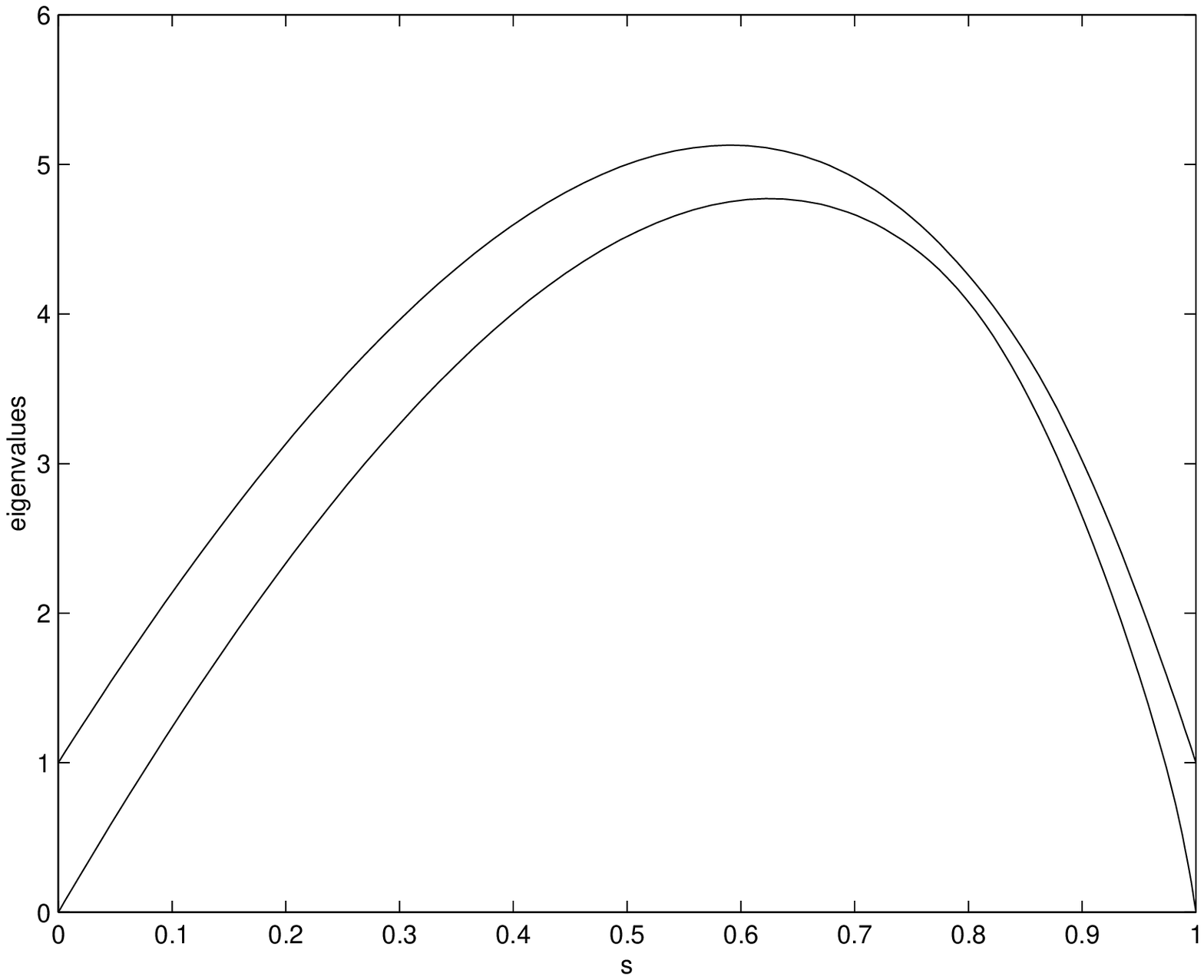 scaled 600}

\numFig{10}{\slshape  The bush of implications; $\log(\gm)$ versus
$\log(n)$ with $n$ ranging from 20 to~120. The straight line indicates that
$\gm \sim n^{-p}$.}{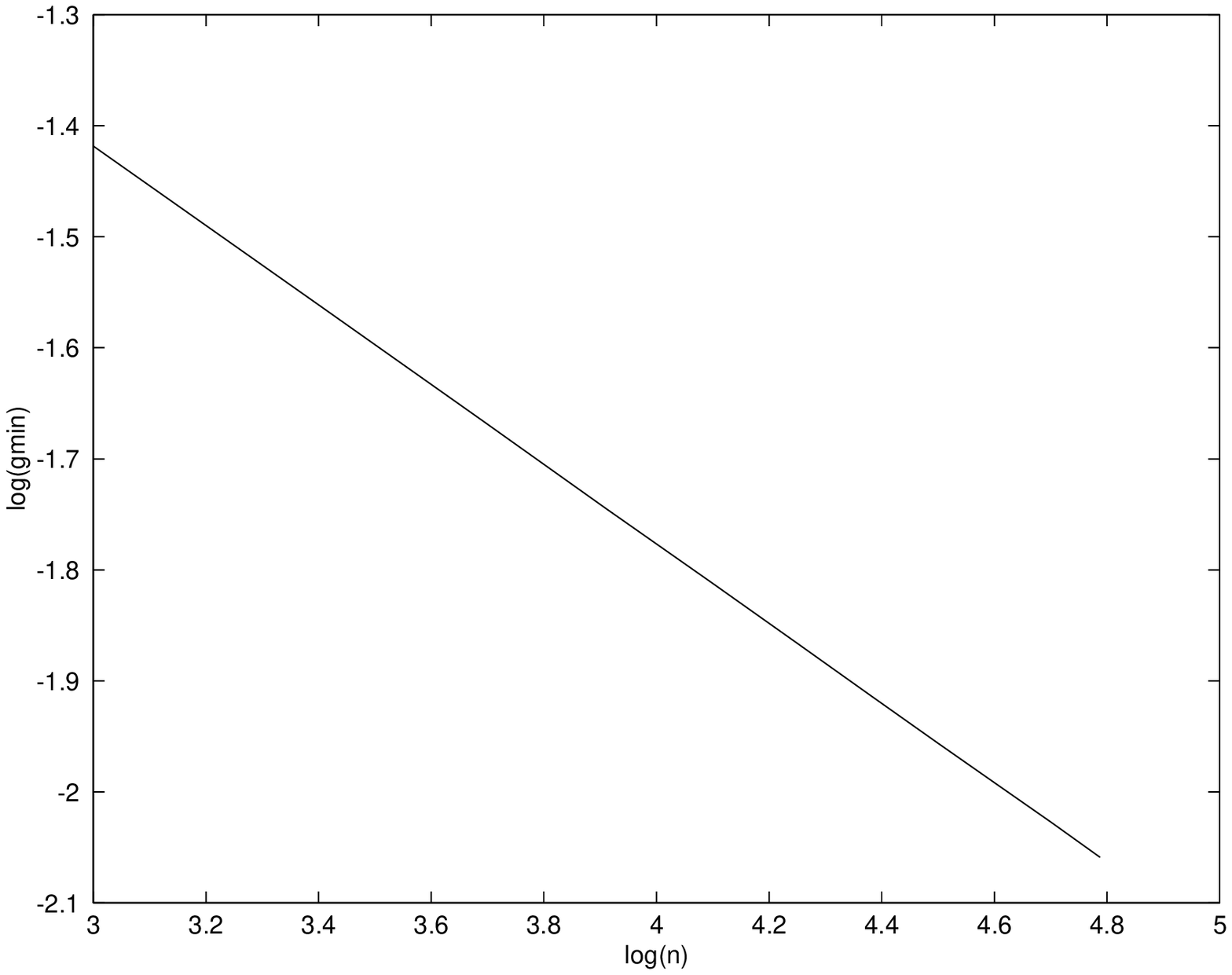 scaled 600}

Given \refEq{3.47} we have numerically evaluated the eigenvalues of  the
$2(n+1)$-dimen\-sional matrix
 with elements
\numEq{4.50}{
\bigl(\bra{z'_0} \bra{m'_z}\bigr) \tilde H(s) \bigl(\ket{z_0} \ket{m_z}\bigr)
} for values of~$n$ in the range from 20 to~120. The two lowest
eigenvalues are shown in
\refFig9 for $n=50$. The gap is clearly visible. In \refFig{10} we plot
$\log(\gm)$ versus
$\log(n)$ and a power law dependence is clearly visible. We conclude that
$\gm\sim n^{-p}$ with $p\simeq \fract38$. For this problem the maximum
eigenvalue of $H_\rB$ is
$2n+1$ and the maximum eigenvalue of $H_\rP$ is $n+1$, so $\cal E$,
which appears in
\refEq{1.8}, at most grows linearly with~$n$. Therefore we have that
with~$T$ of order
$n^{(1+2p)}$ adiabatic evolution is assured. 

We also analyzed adiabatic evolution for the bush of implications using a
different prescription for the initial Hamiltonian. We tried 
\numEq{4.51}{ H'_\rB = \sum_{i=1}^n H_\rB^{(i)} } as opposed to
\refEq{2.21}. This has the effect of replacing the factor of $(n+1)$ in
\refEq{4.47} with a~1. The effect on $\gm$ is dramatic. It now appears to
be exponentially small as a function of~$n$. This means that with the
choice of $H'_\rB$ above, quantum adiabatic evolution fails to solve the
bush of implications in polynomial time. This sensitivity to the distinction
between $H_\rB$ and $H'_\rB$ presumably arises because bit~0 is
involved in $(n+1)$ clauses. This suggests to us that if we restrict attention
to problems where no bit is involved in more than, say, 3~clauses, there
will be no such dramatic difference between using  $H_\rB$ or~$H'_\rB$.

\subsection{Overconstrained 2-SAT} In this section we present another
2-SAT problem consisting entirely of agree and disagree clauses. This time 
every  pair of bits is involved in a clause. We suppose the clauses are
consistent, so there are exactly~2 satisfying assignments, as in
Section~\ref{sec:3.1}. In an $n$-bit instance of this problem, there are
$\left(\begin{smallmatrix}n\\[0.5ex]2\end{smallmatrix}\right)$ clauses,
and obviously the collection of clauses is highly redundant in determining
the satisfying assignments. We chose this example to explore whether this
redundancy could lead to an  extremely small $\gm$. In fact, we will give
numerical evidence that $\gm$ goes like $1/n^p$ for this problem, whose
symmetry simplifies the analysis. 

 As with the problem discussed in Section~\ref{sec:3.1}, at the quantum
level we can restrict our attention to the case of all agree clauses, and we
have
\numEq{3.48}{ H_\rP=\sum_{j<k} H^{jk}_{\rm agree} \ . } Each bit
participates in $(n-1)$ clauses, so when constructing $H_\rB$ using
\refEq{2.21} we take $d_i=n-1$ for all~$i$.  We can write $\tilde H(s)$
explicitly for this problem
\numEq{3.49}{
\tilde H (s) = (1-s) (n-1)\sum^n_{j=1} \half (1-\sigma_x^{(j)}) +  s
\sum_{j<k}  \half (1-\sigma_z^{(j)} \sigma_z^{(k)}) } which in terms of the
total spin operators $S_x$ and $S_z$ is
\numEq{3.50}{
\tilde H(s) = (1-s) (n-1)\Bigl[\frac{n}{2} -S_x\Bigr] +s \Bigl[\frac{n^2}{4} -
S_zS_z
\Bigr]
\ . } As in Section~\ref{sec:3.3}, it is enough to consider the symmetric
states
$\ket{m_z}$. Using \refEq{3.47}, we can find the matrix elements
\numEq{3.51}{
\bra{m'_z}\tilde H(s) \ket{m_z} } and numerically find the eigenvalues of
this $(n+1)\times(n+1)$-dimensional matrix. 

\numFig{11}{\slshape  The two lowest eigenvalues of $\tilde H(s)$,
restricted to the invariant subspace, for overconstrained 2-SAT
 with $n=33$. The visible gap indicates that $\gm$ is not exponentially
small.}{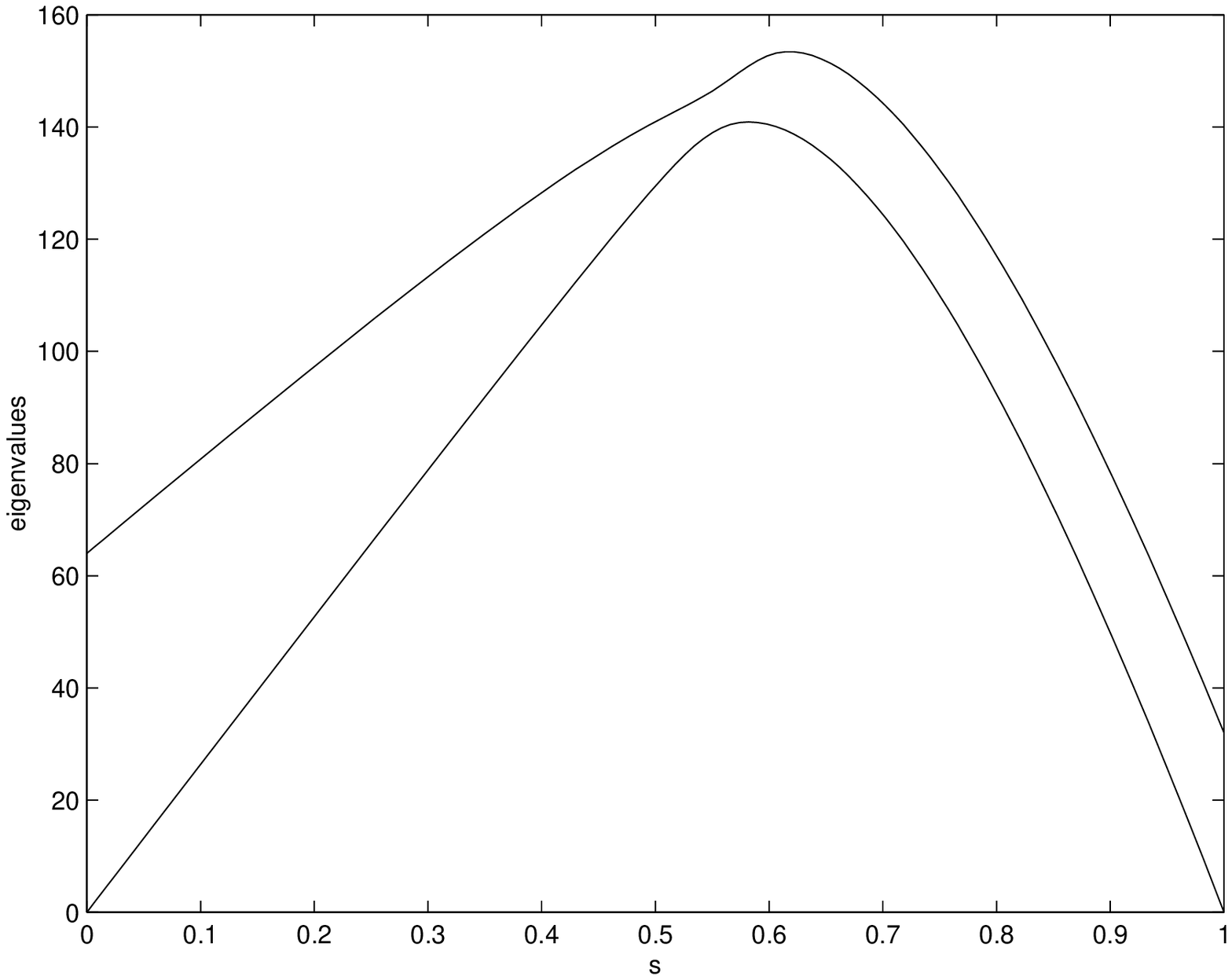 scaled 600}

\numFig{12}{\slshape  Overconstrained 2-SAT; $\log(\gm)$ versus
$\log(n)$ with $n$ ranging from 33 to~203. The straight line indicates that
$\gm \sim n^{p}$.}{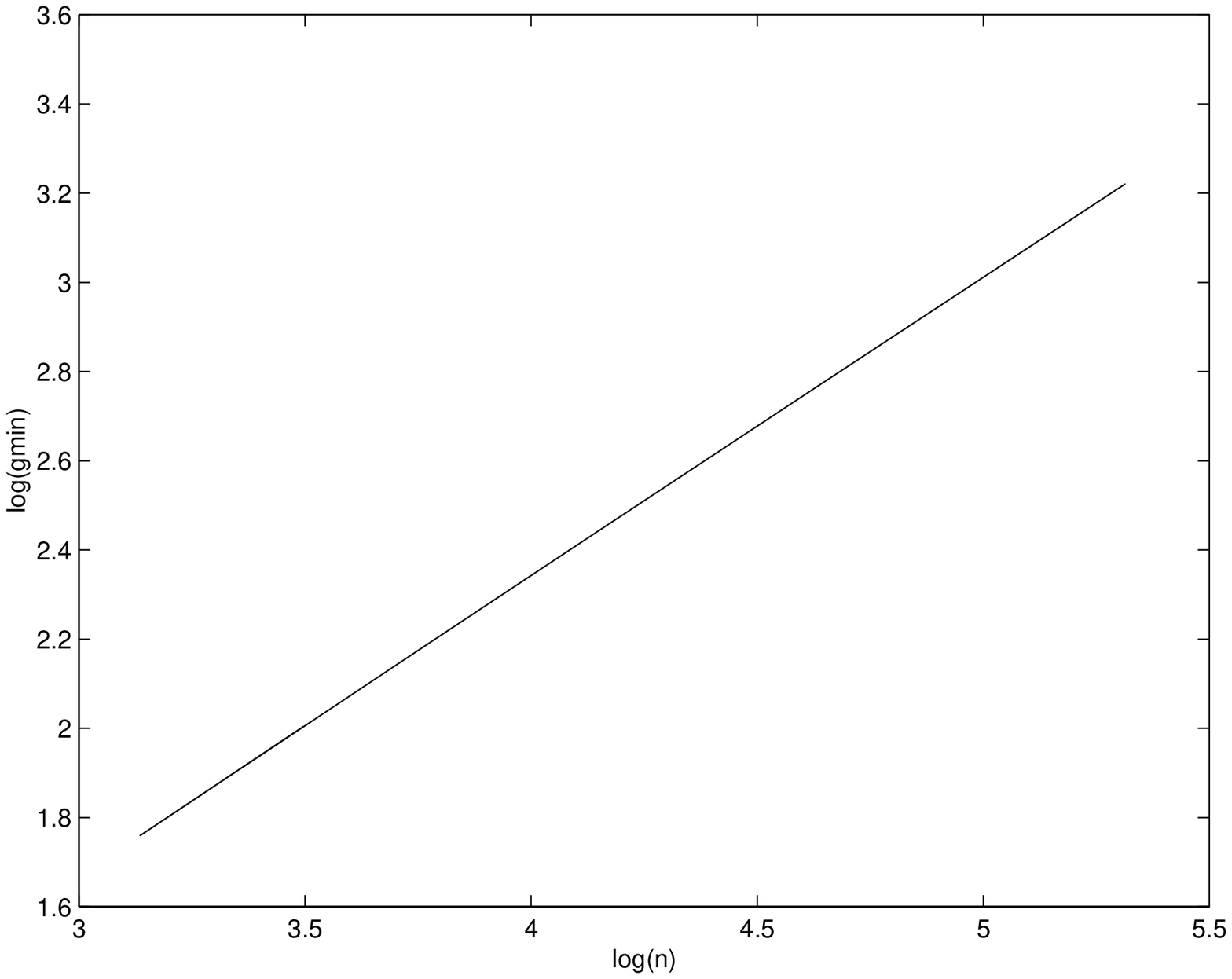 scaled 600}

Actually there are two ground states of $\tilde H(1)$, $\ket{m_z=\fract
n2}$ and 
$\ket{m_z=-\fract n2}$, corresponding to all bits having the value~0 or all
bits having the value~1. The Hamiltonian $\tilde H(s)$ is invariant under
the  operation of negating all bits (in the $z$~basis) as is the initial state
given by \refEq{1.18new}. Therefore we can restrict our attention to
invariant states. In \refFig{11} we show the two lowest invariant states for
33~bits. The gap is clearly visible. ($E_1(0) = 64=2(33-1)$ because the
invariant states all have an even number of~1's in the
$x$-basis.) In
\refFig{12} we  plot $\log(\gm)$ against $\log(n)$. The straight line shows
that
$\gm\sim n^p$ with $p\simeq 0.7$. For this problem the maximum
eigenvalues of $H_\rB$ and $H_\rP$ are both of order $n^2$ so $\cal E$
appearing in
\refEq{1.8} is no larger than~$n^2$.  Adiabatic evolution with $T$ only as
big as
$n^{(2-2p)}$ will succeed in finding the  satisfying assignment for this set
of problems.

\section{The Conventional Quantum Computing Paradigm}\label{sec:4}
The algorithm described in this paper envisages continuous-time evolution
of a quantum system, governed by a smoothly-varying time-dependent
Hamiltonian.  Without further development of quantum computing
hardware, it is not clear whether this is more or less realistic than
conventional quantum algorithms, which are described as sequences of
unitary operators each acting on a small number of qubits.  In any case, our
algorithm can be recast within the conventional quantum computing
paradigm using the technique introduced by Lloyd~\citeRef4.

The Schr\"odinger equation \refEq{1.1} can be rewritten for the unitary
time evolution operator $U(t,t_0)$,
\numEq{4.1}{ i \frac{\rd}{\rd t} U(t,t_0) = H(t) U (t,t_0) } and then
\numEq{4.2}{
\ket{\psi(T)} = U(T,0) \ket{\psi(0)} \ . } To bring our algorithm within the
conventional quantum computing paradigm we need to approximate
$U(T,0)$ by a product of few-qubit unitary operators.  We do this by first
discretizing the interval $[0,T]$ and then applying the Trotter formula at
each discrete time.

The unitary operator $U(T,0)$ can be written as a product of $M$ factors
\numEq{4.3}{ U(T,0) = U(T,T-\Delta ) U(T-\Delta, T-2\Delta )\cdots
U(\Delta,0 ) } where $\Delta=T/M$.  We use the approximation
\numEq{4.4}{ U((\ell+1)\Delta, \ell \Delta ) \simeq e^{-i\Delta H(\ell\Delta)}
} which is valid in \refEq{4.3} if
\numEq{4.5}{
\bigl\|  \Delta H (t_1) - \Delta H (t_2) \bigr\|  \ll \frac{1}{M} \quad
\hbox{for all}
\quad t_1,t_2 \in  [\ell \Delta, (\ell+1) \Delta] \ . } Using \refEq{1.18} this
becomes
\numEq{4.6}{
\Delta \bigl\|  H_\rP - H_\rB \bigr\| \ll 1 \ . } We previously showed (in
the paragraph after Eq.~\refEq{1.20}) that $\|  H_\rP - H_\rB \|$ grows no
faster than the number of clauses, which we always take to be at most
polynomial in $n$.  Thus we conclude that the number of factors 
$M=T/\Delta$ must be of order $T$ times a polynomial in $n$.

Each of the $M$ terms in \refEq{4.3} we approximate as in \refEq{4.4}. 
Now
$H(\ell \Delta)=uH_\rB + vH_\rP$ where $u=1-({\ell\Delta}/{T})$ and
$v= \ell \Delta/T$ are numerical coefficients each of which is between~0
and~1.   To use the Trotter formula
\numEq{4.7}{ e^{-i\Delta H(\ell \Delta)} \simeq (e^{-i\Delta
uH_\rB/K}e^{-i\Delta vH_\rP/K} )^K } for each $\ell$, $\ell=0,1,\dots,M-1$, 
we need $K\gg M \big(1+\Delta\,\|H_\rB\| + \Delta\, \|H_\rP\|\big)^2$. 
Since $\|H_\rB\| $ and 
$\|H_\rP\|$ are at most a small multiple of the number of clauses, we see
that
$K$ need not be larger than $M$ times a polynomial in $n$.

Now \refEq{4.7} is a product of $2K$ terms each of which is $e^{-i\Delta
uH_\rB/K}$ or $e^{-i\Delta vH_\rP/K}$.  From \refEq{2.21} we see that
$H_\rB$ is a sum of $n$ commuting one-bit operators.  Therefore
$e^{-i\Delta uH_\rB/K}$ can be written (exactly) as a product of $n$
one-qubit unitary operators.  The operator $H_\rP$ is a sum of commuting
operators, one for each clause.  Therefore $e^{-i\Delta vH_\rP/K}$ can be
written (exactly) as a product of unitary operators, one for each clause
acting only on the qubits involved in the clause.

All together $U(T,0)$ can be well approximated as a product of unitary
operators each of which acts on a few qubits.  The number of factors in the
product is proportional to $T^2$ times a polynomial in $n$.  Thus if the
required $T$ for adiabatic evolution is polynomial in $n$, so is the number
of few-qubit unitary operators in the associated conventional quantum
computing version of the algorithm.

\section{Outlook}\label{sec:outlook} We have presented a continuous-time
quantum algorithm for solving satisfiability problems, though we are
unable to determine, in general, the required running time.  The
Hamiltonian that governs the system's evolution is constructed directly
from the clauses of the formula. Each clause corresponds to a single term in
the operator sum that is $H(t)$.  We have given several examples of special
cases of the satisfiability problem where our algorithm runs in polynomial
time.  Even though these cases are easily seen to be classically solvable in
polynomial time, our algorithm operates in an entirely different way from
the classical one, and  these examples may provide a small bit of evidence
that our algorithm may run quickly on other, more interesting cases.

\end{document}